\documentclass[11pt]{article}
\usepackage{jheppub}
\usepackage{color}
\usepackage{subfigure}
\usepackage{graphicx}
\newcommand{\I}[2]{I^{(#1)}_{#2}}
\newcommand{\pp}[2]{D^{(#1)}_{#2}}
\newcommand{\ppp}[3]{(D^{(#1)}_{#2})^#3}

\usepackage{comment}

\def\ep{\epsilon}

\newcommand{\hs}[1]{\hspace*{#1 pt}}

\newcommand{\MB}[2]{\hs{-12} \int\limits_{\hs{15}_{ #1 -i \, \infty}}^{\hs{15}^{ #1 +i\, \infty}} \hs{-15} \frac{d #2}{2\pi i}}

\preprint{
SI-HEP-2017-26}
\title{The Sudakov form factor at four loops in maximal super Yang-Mills theory}

\author[a]{Rutger H. Boels,}
\emailAdd{Rutger.Boels@desy.de}
\affiliation[a]{II. Institut f\"ur Theoretische Physik, Universit\"at Hamburg,  \\ Luruper Chaussee 149, D-22761 Hamburg, Germany}
\author[b]{Tobias Huber}
\emailAdd{huber@physik.uni-siegen.de}
\affiliation[b]{Naturwissenschaftlich-Technische Fakult\"at, Universit\"at Siegen, \\ Walter-Flex-Str.~3, 57068 Siegen, Germany}
\author[c]{and Gang Yang}
\emailAdd{yangg@itp.ac.cn}
\affiliation[c]{CAS Key Laboratory of Theoretical Physics, Institute of Theoretical Physics, \\ Chinese Academy of Sciences, Beijing 100190, China}

\keywords{Form factors, cusp anomalous dimension, nonplanar, Casimir scaling}

\abstract{The four-loop Sudakov form factor in maximal super Yang-Mills theory is analysed in detail. It is shown explicitly how to construct a basis of integrals that have a uniformly transcendental expansion in the dimensional regularisation parameter, further elucidating the number-theoretic properties of Feynman integrals. The physical form factor is expressed in this basis for arbitrary colour factor. In the nonplanar sector the required integrals are integrated numerically using a mix of sector-decomposition and Mellin-Barnes representation methods. Both the cusp as well as the collinear anomalous dimension are computed. The results show explicitly the violation of quadratic Casimir scaling at the four-loop order. A thorough analysis concerning the reliability of reported numerical uncertainties is carried out.}

\begin{document}
\maketitle

\section{Introduction}
Gauge theory is the language of the standard model of particle physics. Even more than 50 years after its first modern formulation by Yang and Mills \cite{Yang:1954ek}, it remains a hard task to compute observables even in a perturbative expansion in the coupling constants. Beyond perturbation theory much less is known in general, with the particular exception of those theories that admit a dual description in the strongly coupled sector, such as that provided by the AdS/CFT correspondence \cite{Maldacena:1997re}. This correspondence is by far best understood for the maximally supersymmetric, $\mathcal{N}=4$, Yang-Mills (SYM) theory based on the $SU(N_c)$ gauge group, in 't~Hooft's planar limit \cite{tHooft:1973alw}. In this limit, where $N_c \rightarrow \infty$, remarkable simplifications occur.  A lighthouse result in this direction is the Beisert-Eden-Staudacher equation~\cite{Beisert:2006ez}: this equation describes a certain observable known as the planar lightlike cusp anomalous dimension (CAD) at all values of the coupling in $\mathcal{N}=4$ and ties into integrability ideas. Weak and strong coupling expansions of this anomalous dimension have been matched to independently obtained results, see e.g.~\cite{Gubser:2002tv, Frolov:2002av, Kruczenski:2002fb, Kotikov:2004er, Bern:2006ew, Cachazo:2006az, Roiban:2007dq, Henn:2013wfa}. However, beyond the planar limit much less is known in general despite some very recent progress in~\cite{Alday:2017xua,Aprile:2017bgs}. For the cusp anomalous dimension no nonplanar correction had been computed in any theory until recently the first numerical result at four loops in $\mathcal{N}=4$ was presented by us in \cite{Boels:2017skl}.

Beyond the AdS/CFT correspondence and especially at weak coupling, the $\mathcal{N}=4$ super-Yang Mills theory is also a time-tested sandbox to explore computational ideas, such as those motivated by Witten's twistor string theory \cite{Witten:2003nn}. These have ignited a long-running program to explore the space of on-shell observables, using on-shell methods. This article is a part of this program, aimed at computing the so-called Sudakov form factor in $\mathcal{N}=4$ SYM theory. This form factor can be used to isolate several interesting universal functions that are contained within it. Prime among these is the lightlike cusp anomalous dimension mentioned above. The cusp anomalous dimension plays a central role in the analysis of infra-red (IR) divergences, as first pointed out in  \cite{Korchemsky:1985xj}.  By extrapolating structures found through three loops a general conjecture was formulated in  \cite{Becher:2009qa} that the nonplanar part of the CAD vanishes in any perturbative gauge theory. This became known as  \emph{quadratic Casimir scaling} of the CAD, see e.g. \cite{Korchemsky:1988si, Gardi:2009qi, Dixon:2009gx, Becher:2009qa, Becher:2009kw, Dixon:2009ur, Ahrens:2012qz}. It was noted that the quadratic Casimir scaling may be violated to higher orders of perturbative expansion due to the appearance of higher Casimir operators of the gauge group \cite{Frenkel:1984pz}, see also \cite{Alday:2007mf}. At strong coupling, this scaling is known to break down in $\mathcal{N}=4$ SYM \cite{Armoni:2006ux}. In addition, instanton effects break the scaling \cite{Korchemsky:2017ttd}. Finally,  ref.~\cite{Boels:2017skl} disproved the conjecture in perturbation theory, see also the two recent works \cite{Grozin:2017css,Moch:2017uml} which apply directly to quantum chromodynamics and also report violation of Casimir scaling.

The Sudakov form factor we consider is an observable which involves two on-shell massless states and a gauge invariant operator in the stress tensor multiplet in $\mathcal{N}=4$ SYM,
\begin{equation}
{\cal F} = \int d^4 x \, e^{-i q\cdot x} \langle p_1, p_2 | \mathcal{O}(x) |0 \rangle \,.
\end{equation}
In ${\cal N}=4$ SYM, form factors were first studied thirty years ago in \cite{vanNeerven:1985ja} and revived in the past few years at weak coupling \cite{Brandhuber:2010ad, Bork:2010wf, Brandhuber:2011tv, Bork:2011cj,  Henn:2011by, Gehrmann:2011xn, Brandhuber:2012vm, Bork:2012tt, Engelund:2012re, Johansson:2012zv, Boels:2012ew,  Penante:2014sza, Brandhuber:2014ica, Bork:2014eqa, Bork:2015fla, Frassek:2015rka, Boels:2015yna,  Huang:2016bmv, Koster:2016ebi, Koster:2016loo, Chicherin:2016qsf, Bork:2016hst, He:2016dol, Brandhuber:2016xue,  He:2016jdg, Yang:2016ear, Koster:2016fna, Chicherin:2016ybl, Bork:2017qyh, Meidinger:2017hvm} and at strong coupling \cite{Alday:2007he, Maldacena:2010kp, Gao:2013dza}. There have been interesting recent studies of loop form factors of non-Bogomolnyi--Prasad--Sommerfield (BPS) operators \cite{Wilhelm:2014qua, Nandan:2014oga, Loebbert:2015ova, Caron-Huot:2016cwu, Brandhuber:2016fni, Ahmed:2016vgl, Loebbert:2016xkw, Banerjee:2016kri, Brandhuber:2017bkg}. For reviews, see the theses \cite{Wilhelm:2016izi, Penante:2016ycx}. The present paper is aimed at elucidating the evaluation of the integrals that appear in the four-loop Sudakov form factor, with the expectation that the presented techniques can be applied more widely. 

A key idea in this article is to make transparent the transcendentality properties of the Feynman integrals that make up the Sudakov form factor. It is known quite generally that at fixed orders in the expansion in the dimensional regularisation parameter $\epsilon$ of  Feynman integrals only rational linear combinations of certain constants appear. These constants are known as multiple zeta values (MZV). In principle, also more general constants such as Euler sums can appear, but in the known terms of the Sudakov form factor through to three loops in $\mathcal{N}=4$ SYM, MZVs are sufficient.
MZVs have a property known as transcendental weight which takes integer values. The number of independent MZVs is small for low weight, and a basis for these constants is formed by (see e.g.~\cite{Blumlein:2009cf})
\begin{equation}
\{1\}_0, \{\phantom{v}\}_1, \{\pi^2\}_2, \{\zeta_3\}_3,\{\pi^4\}_4, \{\pi^2 \zeta_3, \zeta_5\}_5, \ldots
\end{equation}
with increasing weight denoted by the subscripts. At fixed order in $\epsilon$ in a generic integral only terms up to a maximal weight appear. This maximal weight increases stepwise with the order of expansion. A special class of integrals is formed by those where only the maximal weight terms appear at each order in the $\epsilon$-expansion. Assigning to $\epsilon$ a transcendental weight $-1$, these integrals have a well-defined overall transcendental weight,
and will be referred to as {\emph{uniformly transcendental}} (UT) integrals.  
The concept of transcendental weight is important as it is observed in many examples that in $\mathcal{N}=4$ SYM (and superstring theory) only terms with maximal weight appear. Although the origin of this is somewhat ill-understood, it at the very least makes for a useful tool. Moreover, a general conjecture \cite{Kotikov:2002ab,Kotikov:2004er} relates the maximal transcendental terms appearing in QCD directly to $\mathcal{N}=4$ SYM for certain quantities. An example of this kind is given by the quark and gluon form factors in QCD~\cite{Baikov:2009bg,Lee:2010cga,Gehrmann:2010ue,Gehrmann:2010tu,vonManteuffel:2015gxa} and the Sudakov form factor in $\mathcal{N}=4$ SYM, where the maximal transcendentality principle was verified through to three loops and for all terms up to transcendental weight eight~\cite{Gehrmann:2011xn}. Examples for two-loop remainders were also found in \cite{Brandhuber:2012vm, Brandhuber:2017bkg}.

For the three-loop form factor in $\mathcal{N}=4$ SYM, an expression in terms of UT integrals was obtained in \cite{Gehrmann:2011xn}. In that case the master integrals were known analytically, facilitating the analysis. In the four-loop case generically the basis of UT integrals was unknown. In this article, it will be shown how to identify UT candidates systematically, and how to write the four-loop Sudakov form factor as a rational linear combination of UT candidate integrals. The result in the nonplanar sector will then be integrated numerically, yielding a large list of new integral results. What is surprising is the empirical observation that obtaining numerical results for UT integrals turns out to be substantially simpler than for generic non-UT integrals in the class under study, even though the integration techniques themselves do not make use of the UT property. The result is combined into the nonplanar cusp and collinear anomalous dimensions at four loops. The result for the cusp anomalous dimension was first announced by us in \cite{Boels:2017skl}, while the result on the collinear anomalous dimension is new. We comment extensively on the numerics below, making use of the UT property to inform the error analysis.

This article is structured as follows: section \ref{sec:rev} contains a review and setup of the problem. In section \ref{sec:UT}, uniformly transcendental integrals are discussed both at the general level as well as for the specific observable under study. Of special interest is a general technique for obtaining candidate-UT integrals. The full form factor is expanded in terms of the UT basis in section \ref{sec:FFinUT}.  In section \ref{sec:numeric} we discuss the numerical integration of the appearing integrals in the nonplanar sector, present our results and perform a thorough analysis of the reported numerical uncertainties. We conclude in section~\ref{sec:conclusion}. The article is supplemented by several appendices. In appendix~\ref{sec:utints} we give explicit results of the UT integrals in the nonplanar sector, while appendix~\ref{app:basis} contains the parametrisation of the integral topologies in terms of loop and external momenta.

\section{Review and setup}\label{sec:rev}

\subsection{Infra-red divergent structure of the form factor in ${\cal N}=4$ SYM}

The perturbative expansion of the Sudakov form factor is fixed by supersymmetry and dimensional analysis as
\begin{equation}
{\cal F} = {\cal F}^{\textrm{tree}} \sum_{l=0}^\infty g^{2 {l}} (-q^2 )^{-l \epsilon } F^{(l)}  \,,
\end{equation}
where $p_1,p_2$ are two on-shell momenta, and $q=(p_1+p_2)$ is off-shell. In dimensional regularisation with $D=4-2 \epsilon$, $F^{(l)}$ is a purely numerical function of gauge group invariants and $\epsilon$. The coupling constant is normalised as $g^2 = \frac{g_{\rm YM}^2N_c}{(4\pi)^2}(4\pi e^{- \gamma_{\text{E}}})^\epsilon$. We consider explicitly the $SU(N_ c)$ gauge group, although our results apply to any Lie group: up to the order considered, there is a one-to-one map from $N_c$ to Casimir invariants, see below. 

The form factor is free of ultraviolet (UV) divergences, since the operator $\mathcal{O}$ in the stress tensor multiplet is protected. On the other hand, there are IR divergences due to soft and collinear singularities from the massless states. Setting $q^2=-1$ and defining the normalised form factor as $F = 1+ \sum_{l=1}^\infty g^{2 {l}} F^{(l)}$, the IR structure is described in the following form \cite{Bern:2005iz} (exponentiation structure of Sudakov form factor in more general theories was original studied in \cite{Mueller:1979ih, Collins:1980ih, Sen:1981sd, Magnea:1990zb})
\begin{align}
\log F = - \sum_{l=1}^\infty g^{2 l} \bigg[ \frac{\gamma_{\textrm{cusp}}^{({l})} }{(2 {l} \epsilon)^2} + \frac{{\cal G}_{\textrm{coll}}^{({l})} }{2 {l} \epsilon} + {\rm Fin}^{(l)} \bigg] + {\mathcal O}(\epsilon) \,, \label{eq:logFF}
\end{align}
where the leading singularity is determined by the cusp anomalous dimension (CAD) $\gamma_{\textrm{cusp}}$, and the sub-leading divergence is related to the so-called collinear anomalous dimension ${\cal G}_{\textrm{coll}}$.\footnote{There are different conventions of defining cusp and collinear anomalous dimensions in the literature. In our convention, the cusp anomalous dimension  $\gamma_{\textrm{cusp}} = \sum_{l}\gamma_{\textrm{cusp}}^{({l})} g^{2l}$ is the same as the function $f(g)$ in \cite{Beisert:2006ez}.}

Besides analysing the IR structure of the form factor, one also has to investigate its colour structure. For a classical Lie-group with Lie-algebra $[ T^{a} ,T^{b} ] = i f^{abc} \, T^c$ and structure constants $f^{abc}$, the quadratic Casimir operators in the fundamental ($F$) and adjoint ($A$) representation are defined via (see e.g.~\cite{vanRitbergen:1998pn})
\begin{align}
[ T^{a} T^{a} ]_{ij} =& C_{F} \delta_{ij} \,, \quad f^{acd} f^{bcd} = C_{A} \delta^{ab} \,,
\end{align}
respectively. The building block of the quartic Casimir invariant $d_{R}^{abcd}d_{R}^{abcd}$ is the fully symmetric tensor
\begin{align}
d_{R}^{abcd} =& \frac{1}{6} {\rm Tr}[ T_R^a T_R^b T_R^c T_R^d + {\text{perms.}(b,c,d)} ] \, ,
\end{align}
where $R = F,A$ denotes the fundamental or adjoint representation, with $ [T_F^a ]_{ij} = [T^a ]_{ij} $ and $ [T_A^a ]_{bc} = -i f^{abc}$. The values of the relevant Casimir invariants in the case of gauge group $SU(N_c)$ read $C_F = N_A/(2N_c)$, $C_A=N_c$, and $d_{A}^{abcd}d_{A}^{abcd}/N_A = N_c^2/24 \, (N_c^2+36)$. Here, $N_A=(N_c^2-1)$ is the number of generators of $SU(N_c)$. The colour structure of the form factor at $l$ loops in ${\cal N}=4$ SYM theory where matter is always in the adjoint representation is simply $(C_A)^l$ up to $l=3$. Starting from four loops, the quartic Casimir invariant arises in addition, and hence in $SU(N_c)$ gauge theory one has, besides the {\emph{planar}} (i.e.\ $N_c^l$ leading-colour) contribution a {\emph{nonplanar}} (i.e.\ $N_c^{l-2}$ subleading-colour) correction. Starting from six loops, additional group invariants appear~\cite{Boels:2012ew}.

The planar form factor has leading divergence $\propto 1/\epsilon^{2l}$ at $l$-loop order. To compute the CAD, this function needs to be expanded down to $\epsilon^{-2}$ at $l$ loops, combined together with higher terms in the Laurent expansion in $\epsilon$ from lower-loop contributions. As mentioned above, the first nonplanar correction starts at four loops, due to the appearance of a quartic Casimir invariant. The nonplanar part of the four-loop form factor takes the following form
\begin{equation}\label{eq:centralrelation}
F^{(4)}_{\textrm{NP}}  = -\frac{\gamma_{\textrm{cusp, NP}}^{(4)}}{(8\epsilon)^2} - \frac{{\cal G}_{\textrm{coll,NP}}^{({4})} }{8 \epsilon} - {\rm Fin}_{\textrm{NP}}^{(4)} + {\mathcal O}(\epsilon) \,.
\end{equation}
In particular, it has only a double pole in $\epsilon$ since, upon taking the logarithm in~(\ref{eq:logFF}), this piece cannot mix with any planar contribution from lower loops. We emphasise that individual integrals that contribute to $F^{(4)}_{\textrm{NP}}$ will typically have the full $1/\epsilon^8$ divergence. The cancellation of these higher-order poles in the final result therefore provides a very strong constraint on as well as a non-trivial consistency check of the computation.

The form factor exhibits a Laurent expansion in the dimensional regularisation parameter $\epsilon$. In this expansion, each term is expected to be a rational-coefficient polynomial of Riemann Zeta values $\zeta_n$, or their multi-index generalizations, $\zeta_{n_1,n_2,...}$, known as multiple zeta values (MZVs)~(see e.g.~\cite{Blumlein:2009cf}). In principle, even more general objects such as Euler sums can appear. However, as mentioned earlier, any analytically known piece of the form factor does not go beyond MZVs. The MZVs have a transcendentality degree which is the sum of their indices, $\sum_{i} n_i$.  Also, the regularisation parameter $\epsilon$ is assigned transcendentality $-1$. In ${\cal N}=4$ SYM, the finite part of the form factor is expected to have (maximal) uniform transcendentality, which at $l$ loops is $2l$, and which suggests that the CAD at $l$ loops is of uniform transcendental weight $2l-2$.
Indeed, the planar CAD at four loops in ${\cal N}=4$ SYM has transcendentality six and was computed as~\cite{Bern:2006ew,Cachazo:2006az, Henn:2013wfa}
 \begin{equation}
 \label{eq:planar-4loop-FF}
(\log F)^{(4)}_{\rm P}  = - \left[ \frac{- 1752 \zeta_6 - 64\zeta_3^2}{(8\epsilon)^2} + \frac{{\cal G}_{\textrm{coll,P}}^{({4})} }{8 \epsilon} \right] + {\mathcal O}(\epsilon^{0}) \,. 
\end{equation}
We will provide strong evidence that also the nonplanar form factor and in particular the CAD are of uniform transcendentality at four loops. A numerical result of the planar four-loop collinear anomalous dimension ${\cal G}_{\textrm{coll,P}}^{({4})}$ was obtained in~\cite{Cachazo:2007ad}. Recently, also the analytic value of this quantity was presented~\cite{Dixon:2017nat}.

\subsection{Integrand and integral relations}

\begin{figure*}[t]
\includegraphics[width=0.99\textwidth]{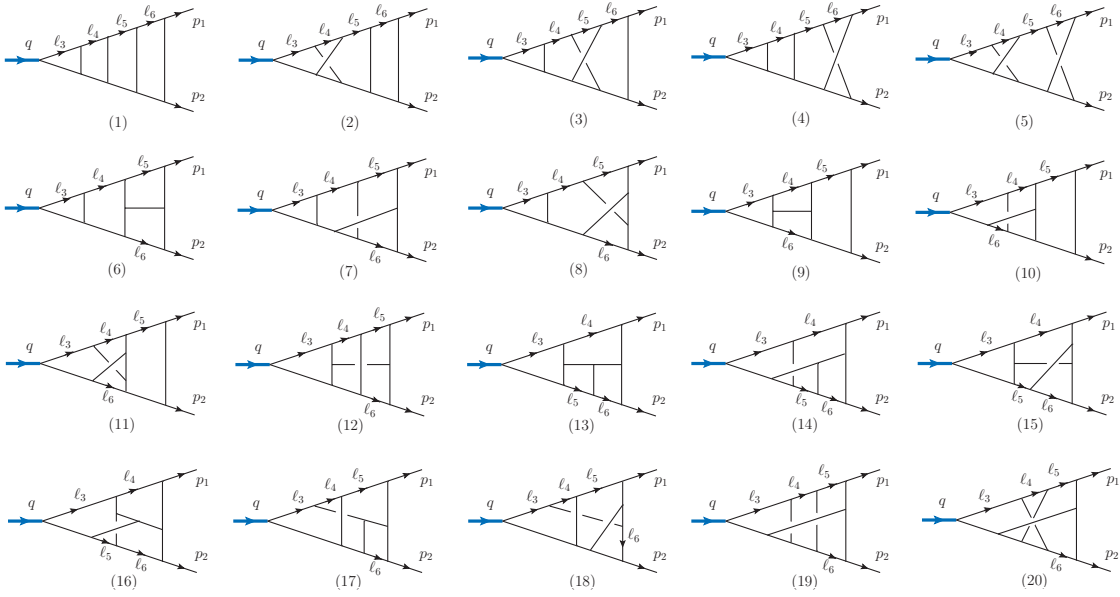}
\caption{\label{fig:Ptops} Integral topologies that contribute only to the planar form factor at four loops.}
\end{figure*}

\begin{figure*}[t]
\includegraphics[width=0.99\textwidth]{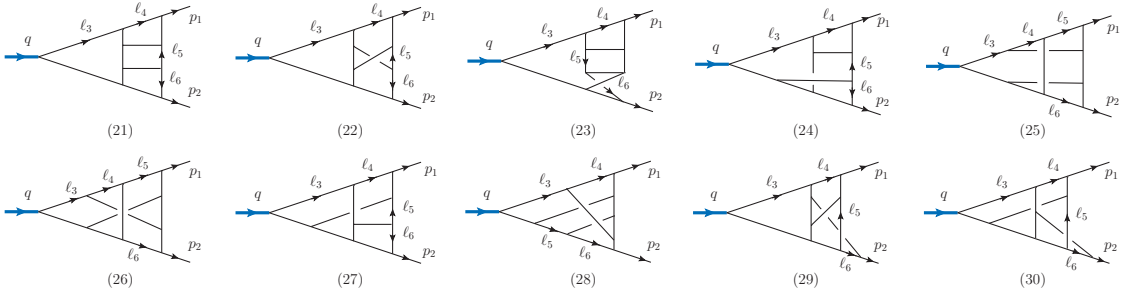}
\caption{\label{fig:NPtops} Sample integral topologies that contribute to the nonplanar form factor at four loops.}
\end{figure*}

\begin{figure}[t]
\centering
\includegraphics[clip,scale=0.4]{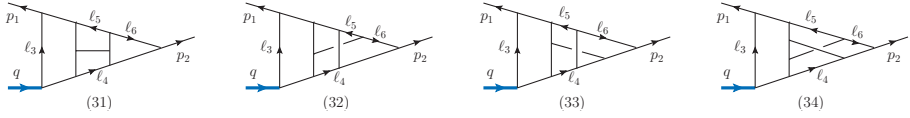}
\caption{Integral topologies that do not have dLog numerators.}
\label{fig:NPtops_NOdlog}
\end{figure}

The full four-loop Sudakov form factor including the nonplanar part in $\mathcal{N}=4$ SYM was obtained as a linear combination of a number of four-loop integrals in~\cite{Boels:2012ew} based on colour-kinematics duality \cite{Bern:2008qj,Bern:2010ue}. Similar five-loop result was also obtained recently in \cite{Yang:2016ear}. For more details on colour-kinematics duality, see e.g. the lecture~\cite{Carrasco:2015iwa}.
The explicit form of the integrals for the problem at hand can be found in~\cite{Boels:2012ew}.  There are $34$ distinct cubic integral topologies, each with $12$ internal lines, that contribute to the four-loop form factor. They are labelled $(1)$\,--\,$(34)$ in~\cite{Boels:2012ew} and we provide them in figures~\ref{fig:Ptops}~--~\ref{fig:NPtops_NOdlog} for convenience and further reference throughout the present paper.

The four-loop integrals take the generic form as
\begin{equation}\label{eq:deffeynmanint}
I = (-q^2)^{2+4\epsilon} e^{4\epsilon\gamma_E} \int {d^D l_1 \over i\pi^{D/2}} \ldots {d^D l_4 \over i\pi^{D/2}} \; \frac{{N}(l_i, p_j)}{ \prod_{k=1}^{12} D_k } \,,
\end{equation}
where $D_i$ are twelve propagators and ${N}(l_i, p_j)$ are dimension-four numerators in terms of Lorentz products of the four independent loop and two independent external on-shell momenta. For each topology, one needs to pick six additional propagators (i.e.\ six irreducible numerators) to form a complete basis, and we label them $D_k, k=13,\ldots, 18$. Such a choice is not unique. Below we use as propagator basis $D_i^{(n)}$, where the superscript $(n)$ indicates the topology, and the subscript $i, i=1,...,18$ refers to the basis given explicitly in appendix \ref{app:basis} (see also Appendix C of \cite{Boels:2015yna}). We define $D_{19}^{(n)} = (p_1+p_2)^2$. Any given numerator can then be represented uniquely in the chosen basis.

A fundamental property of Feynman integrals, as those in equation (\ref{eq:deffeynmanint}), is that they obey integration-by-parts (IBP) identities \cite{Chetyrkin:1981qh, Tkachov:1981wb}, which follow from 
\begin{equation}
\int d^D l_1 \ldots d^Dl_L  \,\, \frac{\partial}{\partial l_i^{\mu}} \, ({\rm integrand}) = 0 \,.
\end{equation}
Working out the left-hand side gives a linear relation between different integrals. By solving linear systems of such equations, a generic Feynman integral can be expressed in terms of a set of basis integrals. This procedure is known as IBP reduction, and the set of basis integrals is also known as the set of master integrals.
The form factor was expressed in terms of a set of master integrals in~\cite{Boels:2015yna} using the Reduze code~\cite{vonManteuffel:2012np}.\footnote{There exist various private and public implementations of IBP reduction, mainly based on Laporta's algorithm \cite{Laporta:2001dd}, such as {\tt AIR} \cite{Anastasiou:2004vj}, {\tt FIRE} \cite{Smirnov:2008iw, Smirnov:2013dia, Smirnov:2014hma} and {\tt Reduze} \cite{vonManteuffel:2012np,2010CoPhC.181.1293S}.  See {\tt LiteRed} \cite{Lee:2012cn, Lee:2013mka} for an alternative approach to IBP reduction.} 
The master integrals, however, have evaded full integration so far due to their overwhelming complexity. In addition, the full IBP reduction generically leads to coefficients that contain higher-order poles in $\epsilon$. This requires to evaluate the master integrals to higher orders in the $\epsilon$ expansion, which further increases the size of the problem. In this paper a different strategy will be used by expanding the form factor in terms of a set of integrals which are each simple enough to integrate and have $\epsilon$-independent prefactors.

\begin{figure}[t]
\centering
\includegraphics[clip,scale=0.5]{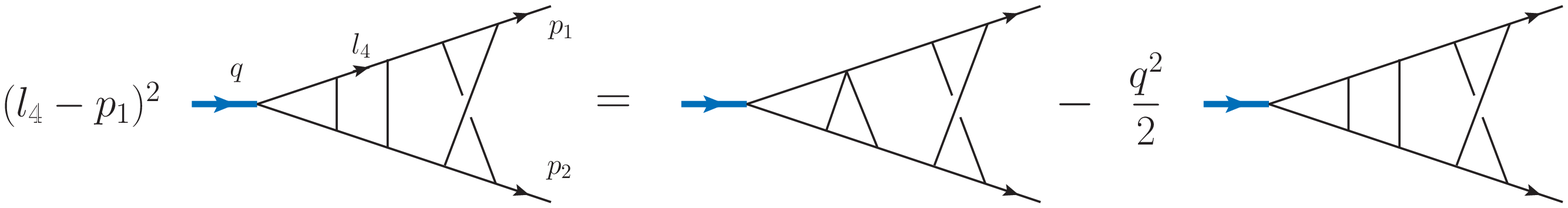}
\caption{Example of rational IBP relations.}
\label{fig:miniIBP}
\end{figure}

A particular subset of the IBP relations turns out to be very useful for our purpose. These are the IBP relations in which the coefficients in front of integrals are pure rational numbers and independent of $\epsilon$. These `rational IBP' relations have been obtained in \cite{Boels:2016bdu} for the form factor presently under study as a subset of the full reduction. 
An example is shown in Fig.\,\ref{fig:miniIBP}. Note that integral relations derived from graph symmetries are a particular subset of the rational IBP relations.

\section{Uniformly transcendental basis}
\label{sec:UT}

A key idea of the present study is to expand the form factor in a set of integrals that all have uniform transcendentality (UT), which will be referred to as UT integrals. Such a representation of the form factor will make manifest the expected maximal transcendentality property of ${\cal N} = 4$ SYM, and has been achieved at three loops in~\cite{Gehrmann:2011xn}. As will be shown in the next section, the UT integrals turn out to be much simpler to integrate numerically compared to generic non-UT integrals of similar complexity, which is crucial for the computation at hand.

We will now turn to the question how to find UT integrals prior to explicitly computing them. There are basically three ways to show whether an integral is UT. 
\begin{itemize}
\item 
A UT integral can be written in the so-called $d$Log form \cite{Arkani-Hamed:2014via, Bern:2014kca}.
\item
The leading singularities, or equivalently, the residues at all poles of a UT integral must always be a constant \cite{Bern:2014kca,Bern:2015ple,Henn:2016men}. This is conjectured to be a necessary and sufficient condition.
\item
A set of UT integral basis can lead to certain simple differential equations \cite{Henn:2013pwa}.
\end{itemize}
The last point regarding differential equations is not directly applicable to the Sudakov form factor at hand since it is a single-scale problem, and thus not `differentiable'. See however \cite{Henn:2013nsa, Henn:2016men} for a work-around by deforming an on-shell leg to be massive, thus creating a two-scale problem. 
Below we illustrate the first two UT properties using a simple one-loop example. Then their application to four-loop form factor integrals will be discussed.

\subsection{Warm up: a one-loop example}

A one-loop UT example is given by the following scalar triangle integral:
\begin{equation}
I_3^{(1)} = (-q^2)^{1+\epsilon} \, e^{\epsilon\gamma_\textrm{E}} \, \int \frac{d^D l}{i\pi^{D/2}} \frac{1}{l^2 (l - p_1)^2 (l + p_2)^2} \,.
\end{equation} 
This is a UT integral as evidenced by the explicit result in the $\epsilon$ expansion
\begin{align}
I_3^{(1)} &= - \, \frac{e^{\epsilon\gamma_\textrm{E}}\Gamma(-\epsilon)^2\Gamma(1+\epsilon)}{\Gamma(1-2\epsilon)}
=  -{1\over \epsilon^2} + {1 \over 2} \zeta_2 + {7 \over 3}\zeta_3 \epsilon  + {47 \over 16} \zeta_4 \epsilon^2  + {\cal O}(\epsilon^3)  \,.
\end{align}
An important interesting point is that, despite that the integral requires regularisation to be well defined, the UT property can be understood in exactly four dimensions at the integrand level. In the following, we consider only the integral in four dimensions as
\begin{equation}
I_3^{(1)} = (-q^2) \int {d^4 l} \frac{1}{l^2 (l - p_1)^2 (l + p_2)^2} \,.
\end{equation}

It is convenient to parametrise the loop momenta such that only scalar integration parameters remain. The four-dimensional loop momentum can be parametrised as
\begin{equation}\label{eq:loopmominparamform}
l = \alpha_1 p_1 + \alpha_2 p_2 + \alpha_3 q_1 + \alpha_4 q_2 \,, 
\end{equation} 
where $p_i=\lambda_i \tilde\lambda_i, i=1,2$ are the external on-shell momenta, and $q_1$, $q_2$ can be chosen as the two complex solutions to 
\begin{equation}
q_i^2 =  q_i \cdot p_j = 0 \quad \forall i,j \quad \textrm{and} \quad q_1\cdot q_2 = -p_1\cdot p_2 \,,
\end{equation}
for example, $q_1 = \lambda_1 \tilde\lambda_2,  q_2 = \lambda_2 \tilde\lambda_1$. The integral in the parametric form is
\begin{equation}
\label{eq:parametric-form}
I_3^{(1)} = \int {d \alpha_1 \, d \alpha_2 \, d \alpha_3 \, d \alpha_4 \over (\alpha_1 \alpha_2 - \alpha_3 \alpha_4)(-\alpha_2 + \alpha_1 \alpha_2 - \alpha_3 \alpha_4)(\alpha_1 + \alpha_1 \alpha_2 - \alpha_3 \alpha_4) } \,.
\end{equation}
This can be written in the following $d$Log form
\begin{equation}
I_3^{(1)} = \int d{\rm Log}(\alpha_1 \alpha_2 - \alpha_3 \alpha_4)\, d{\rm Log}(-\alpha_2 + \alpha_1 \alpha_2 - \alpha_3 \alpha_4) \, d{\rm Log}(\alpha_1 + \alpha_1 \alpha_2 - \alpha_3 \alpha_4)\, d{\rm Log}(\alpha_3) \,,
\end{equation}
which, in terms of momenta, is equivalent to the form
\begin{equation}
I_3^{(1)} = \int d{\rm Log}\big[ l^2 \big] \, d{\rm Log} \big[ (l - p_1)^2 \big] \, d{\rm Log} \big[ (l + p_2)^2 \big] \, d{\rm Log} \big[ -2 (l \cdot q_2) \big] \,.
\end{equation}
The existence of the $d$Log representation implies that the integral is UT.

As mentioned above, an alternative way to prove UT property is to consider the leading singularity. In the parametric form like \eqref{eq:parametric-form}, this is equivalent to check the residues at all poles of the integral: the constant leading singularity property translates to the \emph{simple pole} condition for all parameters. Let us explain this in more detail. To check the simple pole condition, one needs to pick up a certain order of the parameters to take the residue. Consider the one-loop example, we can first take residue for $\alpha_1$ at the pole of the first propagator
\begin{equation}
\textrm{Residue at pole} \ \alpha_1 = {\alpha_3 \alpha_4 \over \alpha_2} \quad \rightarrow \quad \int {\rm d} \alpha_2 {\rm d} \alpha_3 {\rm d} \alpha_4 {1\over \alpha_2 (-\alpha_2 + \alpha_2^2 + \alpha_3 \alpha_4)}  \,.
\end{equation} 
Next, we take the residue for $\alpha_2$ at pole $0$
\begin{equation}
\textrm{Further residue at pole} \ \alpha_2=0 \quad \rightarrow \quad \int {\rm d} \alpha_3 {\rm d} \alpha_4 {1\over \alpha_3 \alpha_4 }  \,.
\end{equation} 
The remaining parameters obviously have only simple poles and the final residue is a constant. 
One needs to check all different orders of taking residues, and in all occurring poles. In any intermediate step, after taking a residue in a particular parameter, if one encounters other than a simple pole in a remaining parameter, the integral is not UT. 

We would like to emphasise that the simple pole requirement should also apply to poles at infinity. To be more concrete, consider following simple examples.
For the integral
\begin{align}
\int d\alpha_1 d\alpha_2 {\alpha_2\over \alpha_1 (1-\alpha_2)^2} \,,
\end{align}
there is a double pole for $\alpha_2$ at $1$, thus it is not UT.
As for another example,
\begin{align}
\int d\alpha_1 d\alpha_2 {1\over \alpha_1} \quad {\rm or} \quad \int d\alpha_1 d\alpha_2 {\alpha_2\over \alpha_1 (1-\alpha_2)}
\end{align}
both have a double pole for $\alpha_2$ at infinity, so they are not UT either.\footnote{We would also like to point that for amplitudes in ${\cal N}=4$ SYM as studied in \cite{Bern:2014kca,Bern:2015ple},  an additional requirement is imposed: here even a simple pole at infinity is not allowed. In that case, this is closely related to the hidden dual conformal symmetry of amplitudes \cite{Bern:2014kca,Bern:2015ple}. For the form factor, we must allow simple pole at infinity.}

The condition that only simple poles are allowed is related to the required existence of a $d$Log form where only a logarithmic singularity is allowed. However, it does not require to find the explicit transformation to the $d$Log form which can be very complicated in general. This simple-pole condition provides an essential constraint for the construction of UT integrals below. A related strategy was also used in \cite{Bern:2014kca,Bern:2015ple,Henn:2016men}.

\subsection{Systematic construction}\label{subsec:systematiconstruc}

The aforementioned condition of simple poles is used here both to construct and as well as to check UT integral candidates. Given a four-loop form factor integral in four dimensions, there are $16$ integration parameters, so in principle there are $16! \sim 2\times 10^{13}$ different orders in which the residues can be taken. Practically therefore, the simple-pole condition is verified by choosing a large number of random orders of taking residues. A non-UT integral typically fails the UT test well within a few hundred of such random checks.

This UT test strategy can be used to constrain the space of potential UT integrals when combined with an Ansatz for the numerator. For the four-loop form factor integrals, one can start with a linear Ansatz of mass dimension four numerators of a given topology. We then perform the above described residue tests. The requirement of absence of higher order poles provides linear constraints on the set of coefficients in the Ansatz by computing the residues of the higher order poles. Solving these linear constraints then yields a smaller Ansatz, and the process is repeated. For speed, it is better to first identify a sequence of residues leading to a higher pole by choosing the Ansatz coefficients to be random integer numbers. This sequence can then be used to derive the analytic constraint on the full Ansatz. 
Below we provide more technical details. 

A full four-loop topology contains 12 lines (i.e. propagators). For many topologies one can simply ask the following question: which sets of 10- and 11-line integrals can be added to a given topology in the four-loop form factor such that the sum is UT? Suppose such a linear combination exists. Then it is obvious this is likely not unique: adding any linear combination of 10- or 11-line UT integrals will satisfy the same constraint. To find a basis for all these UT integrals, we take the form factor numerators $N_{\rm ff}$ as they appear in the $\mathcal{N}=4$ theory as input and add to this the set of all $10$- and $11$-line integrals (of which there are $162$). In this case the initial Ansatz looks like
\begin{equation}\label{eq:add11linerstoFF}
N_{\rm ansatz} = a_0 N_{\rm ff} + \sum_{j=1}^{12} D_j \bigg(\sum_{k=13}^{19} a_{j,k}D_k \bigg) + \sum_{1\leq j \leq k \leq 12} b_{j,k} D_j D_k \,,
\end{equation}
where $D_{19}:=q^2$ and the $D$'s are the propagators as given in appendix \ref{app:basis}. Inserting the parametrisation \eqref{eq:loopmominparamform} for each of the four-loop momenta gives a rational expression of $16$ $\alpha$-type parameters. Now one needs to identify a sequence of residues yielding a double or higher pole in the $\alpha$ parameters. Demanding that the pole becomes a simple one yields at least one constraint equation for the $162+1$ parameters $\{a_0, a_{j,k}, b_{j,k}\}$. Explicitly solving these linear constraint equations gives a smaller Ansatz. Now one repeats by again trying to find a sequence of residues that will yield additional constraints. After a number of iterations for the integrals in the case at hand, one has obtained a set consisting of one integral containing the $12$-line parts of the form factor contribution and other integrals which contain at most $11$ lines. These are a set of UT candidate (UTC) integrals. 

One can also ask the question which UT integrals with unit exponent propagators exist in a given topology, worrying later about expressing the $\mathcal{N}=4$ form factor in terms of these. To answer this question one chooses a more general initial Ansatz such as
\begin{equation}
N_{\rm ansatz} = \sum_{1\leq j \leq k \leq 19} b_{j,k} D_j D_k \,.
\end{equation}
Here the simple pole condition will provide a set of linear equations of $190$ parameters $\{b_{j,k}\}$. The end-result for this wider initial ansatz will be a set which contains all possible UT candidate integrals in a given topology (with unit exponents for the propagators).  

If, after deriving constraints with a certain number of random checks and no new further constraints are found in typically a few hundred more random pole checks, the remaining Ansatz contains a set of good UT candidates. 

The choice of initial Ansatz is dictated to a large part by practical ease of subsequent numerical integration. For many public codes, the numerator of integrals is in general preferred to be a product of two factors, each quadratic in momenta. If a single such integral is to contain the full $12$-line parts of a particular integral topology $k$, a necessary but not sufficient condition is to check that the irreducible numerators of a given integral form a product form separately. Concretely, one sets all propagators of this topology to zero, and verifies if a product form emerges for the irreducible numerators. In our chosen set of expressions, the propagators of a topology are always the first $12$ entries (see Appendix \ref{app:basis}), so to check is: 
\begin{equation}\label{eq:prodformexist}
\textrm{UTC} |_{\textrm{top}_{k}\left(D^{(k)}_i = 0\quad  \forall 1\leq i \leq 12 \right)} \stackrel{?}{=} \textrm{product form} \, ,
\end{equation}
where the product form is a quadratic function of $D_{i}, i=13,...,19$. This condition is satisfied for all topologies in the four-loop form factor under study, except for topologies $(12)$, $(17)$, $(19)$, and $(26)$. Note this condition is independent of the exact choice of propagator basis. If this condition is satisfied, then the smaller Ansatz approach of form factor integral plus 10- and 11-liners has a chance of sufficing. This  is usually much quicker and more transparent. If the 12-line parts do not have a product form, the larger Ansatz must be used. Examples of both possibilities are, for instance, topology $(19)$
\begin{equation}
\textrm{UTC} |_{\textrm{top}_{19} \left(D^{(19)}_i = 0\quad  \forall 1\leq i \leq 12 \right)} = -D^{(19)}_{14}  D^{(19)}_{16} - D^{(19)}_{13} D^{(19)}_{19} \,,
\end{equation}
which does not have a product form, and topology $(23)$
\begin{equation}\label{eq:top23prodform}
\textrm{UTC} |_{\textrm{top}_{23} \left(D^{(23)}_i = 0\quad  \forall 1\leq i \leq 12 \right)} = (D^{(23)}_{13}  + D^{(23)}_{19})^2 \, ,
\end{equation}
which does. 

From a generic set of UT candidates $\textrm{UTC}_i$, the product form can be found by solving the following equation
\begin{equation}\label{eq:prodformans}
\sum_i \lambda_i \ \textrm{UTC}_i = \Big(\sum_j  \alpha_j D_j \Big) \Big(\sum_k \beta_k D_k \Big) \, ,
\end{equation}
for non-trivial parameters $\lambda$, $\alpha$ and $\beta$ which are rational numbers. This is a quadratic set of equations, obtained by matching coefficients of products of $D$'s. Since we are interested in integrals that can be used to express the form factor in, more constraints can be added to the problem for specific purposes. For instance, the constraint can be added that the twelve-line parts match known form factor numerator contribution in the topology under study,
\begin{equation}
\Big(\sum_i \lambda_i \ \textrm{UTC}_i  - \textrm{FF}\Big)_{\textrm{12-line parts}} = 0 \qquad\quad \textrm{as possible constraint.}
\end{equation}
Note this constraint only makes sense in a topology where the form factor has a product form on the left hand side of equation \eqref{eq:prodformexist}. Alternatively, one can simply demand one specific coefficient to be unity,
\begin{equation}
\Big(\sum_i \lambda_i \ \textrm{UTC}_i\Big)_{D_j D_k \ \textrm{coefficient}} = 1 \qquad\quad \textrm{as possible constraint.}
\end{equation}
This in particular avoids finding trivial solutions to the general problem in equation \eqref{eq:prodformans} ($\lambda_i =\alpha_j= \beta_k =0$). This constraint is particularly useful when looking for very general solutions to the quadratic problem, matching only to some terms appearing in the form factor. Finally, one can add manifest graph symmetry constraints on the UT candidates: this we did in almost all cases. Which constraint to use in a particular situation depends on the generality of the solution sought for. 

Having set up the quadratic problem \eqref{eq:prodformans}, the first step is to solve the linear sub-problem for $\lambda$. Then, one can impose graph symmetry patterns on the product form. The remaining set of quadratic equations can be analysed completely, or a particular solution can be guessed by computer algebra.\footnote{In {\tt Mathematica}, these options are represented by the commands {\tt Reduce} and {\tt FindInstance}, respectively.} In several cases, it can be shown that no solution to a given problem exists. In these cases, after exhausting all options, one can widen the Ansatz in equation~\eqref{eq:prodformans} by adding a linear combination of ten-line integrals (which are expected to be simple to integrate). These cases can be  clearly seen in the results in section \ref{sec:FFinUT}, e.g.~\eqref{I_25-5}\,--\,\eqref{I_26-7}. Also, sometimes residual parameter-containing solutions to the product-form problem are obtained. In these cases educated guesses were employed, aimed at as parametrically simple as possible integrals.

The result is a list of product-form UT candidates for each topology. The ones listed in this article have all individually been checked to pass at least $10,000$ simple residue checks, giving ample evidence for their uniform transcendentality. As will be discussed later, checking a set of found integrals individually also serves as a useful cross-check on computational errors. 

\subsection{$d$Log forms}

Writing a four-loop integral in $d$Log form will give a direct proof of UT property. However, the construction of a $d$Log form for a generic four-loop form factor integral is a difficult task, and hence this method is more suitable to {\emph{show}} the UT property of a given integral rather than to {\emph{derive}} a UT numerator.

\begin{figure}[t]
\centering
\includegraphics[clip,scale=0.5]{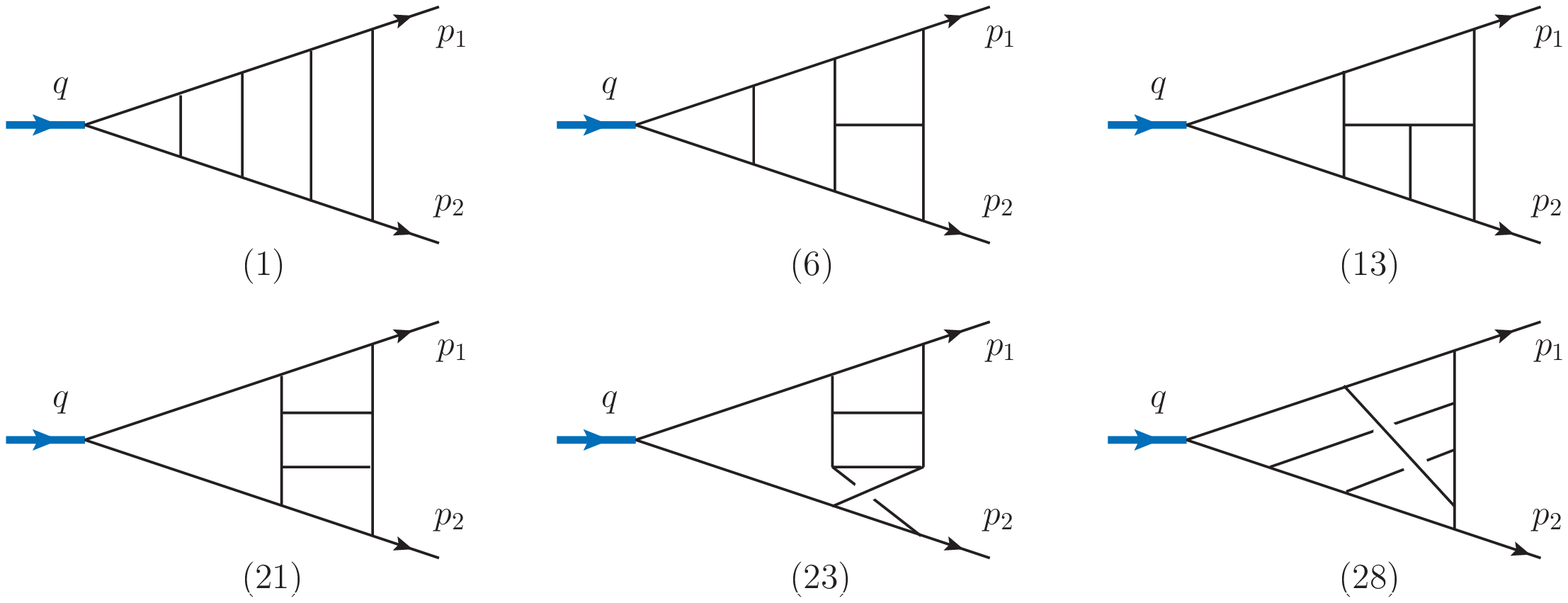}
\caption{Topologies for which it is straightforward to construct a $d$Log form.}
\label{fig:dLogFigs}
\end{figure}

A useful strategy to construct a $d$Log form is loop by loop  \cite{Bern:2014kca,Bern:2015ple}. With proper numerators, all one-loop triangle and box integrals can be written explicitly in $d$Log forms. For example, the three-mass box is known to have a $d$Log form (see e.g.~\cite{Bern:2014kca}, $k_1$ is massless, $K_2$ and $K_4$ are massive)
\begin{align}
\int d^4 \ell {N_\textrm{3m} \over \ell^2 (\ell - k_1)^2 (\ell-k_1-K_2)^2 (\ell +K_4)^2} \,, 
\end{align}
with given numerator
\begin{align}
N_\textrm{3m} = (k_1+K_2)^2 (k_1+K_4)^2 - K_2^2 K_4^2\,,
\end{align}
which is the Jacobian of the quadruple cut of the box, such that the leading singularity is a kinematics-independent constant. So when there is a three-mass sub-box in the four-loop integral, one can write this sub-box in a $d$Log form, and the remaining integral is a three-loop integral involving a new propagator $1/N_\textrm{3m}$. In some topologies, such a procedure can be done recursively loop by loop, so that the full integral can be written explicitly in the $d$Log form. This normally happens when the topology involves at least one box with at least one massless leg, and has some ladder structure.\footnote{It is also possible to write a $d$Log form for four-mass box and three-mass triangle integrals, with numerators in a square-root form. This makes it difficult to find a $d$Log form for the remaining part, since it introduces a square-root propagator. It would be interesting to see if there is a systematic way to solve such cases.} Such cases include topology $(1)$, $(6)$, $(13)$, $(21)$, $(23)$, $(28)$, as shown in  figure~\ref{fig:dLogFigs}, whose  $d$Log  numerators are given, respectively, by
\begin{align}
\big\{ & (q^2)^2,\ (l_4 - p_1)^2 q^2, \ (\ell_3 - p_1)^2(q^2- 2\ell_4\cdot p_2) - (\ell_4 - p_1)^2 (q^2- 2\ell_3\cdot p_2),  \\
& [(l_3 - p_1)^2]^2, \  [(l_3 - p_1)^2]^2, \ (\ell_3 - \ell_4- p_2)^2 \, (\ell_3-p_1)^2\big\} \,. \nonumber
\end{align}

\section{Full form factor in UT basis}
\label{sec:FFinUT}
Finding an expansion of the full form factor in terms of generic UT candidate integrals can be obtained by relatively straightforward linear algebra techniques. In addition, we discussed above how to find product-form numerators for candidate UT integrals. Combining the two involves quite a wealth of choices that can be made in intermediate steps. For the nonplanar form factor, we first found a linear combination of 12-line UT candidates which satisfies the requirement that the difference to the full result contained at most 11-line integrals. Combining the remaining expression into UT candidates in the nonplanar sector was then a relatively easy task. 
In the planar sector, it turned out that more work was required.  An obscuring factor is the existence of many relations between different integrals from the rational IBP relations. 
A choice that works is given below.
This choice was driven by the attempt to find as simple expressions as possible and to express the end-result in as small a number of integrals as possible. This includes both aiming at graph-symmetric expressions as well as trying to find an expansion involving only small integer or half-integer expansion coefficients. This necessarily involves some heuristics. It would be very interesting to find concise target integral expressions more easily, ideally driven by integration convenience or accuracy, but this would lead us beyond the scope of this work.

One important result that follows is that both the planar as well as the nonplanar sector of the form factor can be expressed in terms of rational (i.e.\ $\epsilon$-independent) linear combinations of UT integral candidates. We regard this as strong evidence for the maximal transcendentality of the form factor. By extension, this implies maximal transcendentality for the cusp and collinear anomalous dimensions at the four-loop order in maximal SYM theory, both in the planar and nonplanar sectors. Moreover, the smallness of the expansion coefficients clearly suggests this expansion is natural. In the nonplanar sector we have checked explicitly that the form factor integrals found originally in \cite{Boels:2012ew} when taken as complete topologies can only be expressed in terms of UT integrals in one unique combination of the 14 topologies: the one in which they appear. This provides a cross-check on the symmetry and colour factors.

\subsection{UT integrals for the nonplanar form factor}
\label{sec:NPUT}

Below we list $23$ UT integrals $I^{(n)}_{1\,-\,23}$ that combine into the nonplanar form factor. The superscript $(n)$ denotes the twelve propagators from topology $(n)$ in Fig.~\ref{fig:NPtops}. In this notation, we only have to list the numerator of each integral. Moreover, each integral $I^{(n_i)}_{i}$ gets multiplied by a rational pre-factor $c_i$ according to 
\begin{equation}
\vec c = \{1/2, 1/2, 1/2, -1, 1/4, -1/4, -1/4, 2, 1, 4, 1, 1, -1/2, 1, 1, 1, 1, 1, 1, 1, -1, 1/4, 1/2\} \,.
\end{equation} 
The nonplanar form factor is then obtained as
\begin{equation}
\label{eq:FF_NP}
F^{(4)}_{\textrm{NP}} = {48 \over N_c^2} \sum_{i=1}^{23} \, c_i \, I^{(n_i)}_{i} \,,
\end{equation}
where the prefactor $48/N_c^2 = 2 \times 24/N_c^2$ is the normalisation stemming from the permutational sum of external legs and the colour factor~\cite{Boels:2012ew}, and the UT integrals are 
\allowdisplaybreaks[1]
\begin{align}
I^{(21)}_{1}   &=[(\ell_3-p_1)^2]^2\\[0.2em]
I^{(22)}_{2}   &=-(\ell_3-p_1)^2 \, [\ell_4^2+\ell_6^2-\ell_3^2+(\ell_3-\ell_4+p_1)^2  +(\ell_3-\ell_6-p_1)^2]\\[0.2em]
I^{(23)}_{3}   &=[(\ell_3-p_1)^2]^2\\[0.2em]
I^{(24)}_{4}   &=(\ell_3-p_1)^2 \, [(q-\ell_3-\ell_5)^2 + (\ell_5+p_2)^2 ]\\[0.2em]
I^{(25)}_{5}   &=\left[(p_1 - \ell_5)^2+2 (\ell_4-\ell_5)^2+(\ell_3-\ell_4)^2-(\ell_3-\ell_5)^2 -(p_1-\ell_4)^2 \right]^2 \nonumber \\[0.2em]
               & \quad  -4\, (\ell_4-\ell_5)^2 \, (p_1-\ell_3+\ell_4-\ell_5)^2 \label{I_25-5} \\[0.2em]
I^{(26)}_{6}   &=[(\ell_3-\ell_4-\ell_5)^2-(\ell_3-\ell_4-p_1)^2-(\ell_6-p_2)^2-\ell_5^2]  [\ell_5^2-\ell_4^2-\ell_6^2+(\ell_4-\ell_6)^2] \nonumber \\[0.2em] 
               & \quad  +4\, \ell_5^2 \, (\ell_6-p_2)^2 + (\ell_4-\ell_5)^2 \, (\ell_3-\ell_4+\ell_6-p_2)^2 \\[0.2em]
I^{(26)}_{7}   &=4\, [(\ell_4-\ell_5) (\ell_3-\ell_4+\ell_5-p_1)]   [(\ell_4-\ell_6) (\ell_3-\ell_4+\ell_6-p_2)] \nonumber \\[0.2em]
	       & \quad - 4\, (\ell_4-\ell_5)^2 \, (\ell_3-\ell_4+\ell_6-p_2)^2 - (\ell_3-\ell_4)^2 \, (\ell_5+\ell_6-\ell_4)^2 \nonumber \\[0.2em]
               &  \quad  - \ell_6^2 \, (\ell_5-p_1)^2 - \ell_5^2 \, (\ell_6-p_2)^2  - \ell_4^2 \, (\ell_3-\ell_4+\ell_5+\ell_6-q)^2 \label{I_26-7}\\[0.2em]
I^{(27)}_{8}   &=\frac{1}{2} \left[\ell_3^2 - \ell_4^2 - (\ell_4-\ell_3-p_1)^2\right]   \left[(\ell_3-\ell_4-\ell_5)^2 + (\ell_5+p_2)^2\right] \\[0.2em]
I^{(28)}_{9}   &=(\ell_3 - \ell_4 - p_2)^2 \, \left[ (\ell_3-\ell_4)^2 - (\ell_3-p_1)^2\right]\\[0.2em]
I^{(29)}_{10}  &=\frac{1}{2} \left[\ell_3^2 - \ell_4^2 - (\ell_4-\ell_3-p_1)^2\right] \, \left[\ell_6 \cdot (\ell_6 - \ell_4 + \ell_3 - p_2)\right]\\[0.2em]
I^{(30)}_{11}  &= (\ell_3-\ell_4-p_2)^2 [(p_1-\ell_4)^2+(\ell_3-\ell_4)^2-(\ell_3-p_1)^2] \\[0.2em]
I^{(27)}_{12}  &= \frac{1}{2} \, (\ell_3-\ell_4)^2 \, \left[2 \, (\ell_4-p_2)^2 + (\ell_6-p_1)^2 - \ell_4^2 + \ell_5^2 - (\ell_4-\ell_6)^2 + 2 \, (p_1+p_2)^2\right] \\[0.2em]
I^{(28)}_{13}  &= \frac{1}{2} \, (\ell_3-\ell_4)^2 \, \left[2 \, (\ell_3-\ell_4-p_2)^2 + (\ell_6-p_1)^2 + \ell_4^2   - (\ell_4-\ell_6)^2 \right] \\[0.2em]
I^{(29)}_{14}  &= (\ell_4-p_1)^2 \, \left[(\ell_3-\ell_4+\ell_6)^2 + (\ell_6-p_2)^2 - \ell_6^2\right] \\[0.2em]
I^{(29)}_{15}  &= \frac{1}{2} \, (\ell_3-p_1-p_2)^2 \, \left[(\ell_4-\ell_6)^2 - (\ell_4-p_2)^2  - (\ell_6-p_1)^2 - (p_1+p_2)^2 \right]\\[0.2em]
I^{(30)}_{16}  &= (\ell_3-p_1-p_2)^2 \, (\ell_5+p_2)^2 \\[0.2em]
I^{(30)}_{17}  &= \frac{1}{2} \, (\ell_4-p_1)^2 \, \left[2\, (\ell_5+p_2)^2 - (\ell_5+p_2+\ell_4-\ell_3)^2 \right] \\[0.2em]
I^{(30)}_{18}  &= \frac{1}{2} \, (\ell_3-\ell_4)^2 \, \left[2\, (\ell_6-\ell_4+p_1)^2 - 3 \, \ell_6^2 \right] \\[0.2em]
I^{(22)}_{19}  &= (\ell_3-\ell_4)^2 \,  (p_1-\ell_3+\ell_6)^2 \\[0.2em]
I^{(22)}_{20}  &= \ell_6^2 \,  (p_1-\ell_4)^2 \\[0.2em]
I^{(24)}_{21}  &= (p_1 - \ell_3-\ell_5)^2 \,  (\ell_3-p_1-p_2)^2 \\[0.2em]
I^{(24)}_{22}  &= \ell_5^2 \,  (\ell_3-p_1-p_2)^2 \\[0.2em]
I^{(28)}_{23}  &= (\ell_4 - p_1)^2 \,  (\ell_3-\ell_4+\ell_5-p_2)^2 \, . 
\end{align}
We note that integrals $I_{1\,-\,11}$, $I_{12\,-\,18}$, and $I_{19\,-\,23}$, are 12-, 11-, and 10-line integrals, respectively.

The integral $I^{(25)}_{5}$ in topology $(25)$ is the only one which does not carry the symmetry of the topology explicitly. This was done to arrive at a simpler form to integrate. In general topologies $(25)$ and $(26)$ are the hardest topologies to find UT integrals which are reasonably compact. Note that topologies $(31)$ through $(34)$ do not appear: there are no UT candidate integrals at all in these topologies.

\subsection{UT integrals for the planar form factor}
\label{sec:PLUT}

Similar to the nonplanar part, we also provide an expansion of the planar form factor in terms of $32$ UT integrals $I^{(n)}_{{\rm p},1\,-\,32}$. 
To distinguish from the nonplanar integrals, we add `p' in subscription to denote it is for the planar form factor.
Each integral $I^{(n_i)}_{{\rm p},i}$ gets supplemented by a rational pre-factor $c_{{\rm p},i}$ according to 
\begin{align}
\vec c_{\rm p} = 
& \big\{8, 2, -2, 2, 1/2, 2, 4, 2, -2, 1, 1, 2, 2, -2, 2, -2, 1, 1, 1/2, 2, 
2, 4, -2, -1, 4, -1, -2, -2, \nonumber\\
& \   -1, -1, 1, -1/2 \big\} \,. 
\end{align} 
The planar form factor is then obtained as 
\begin{equation}
\label{eq:FF_PL}
F^{(4)}_{\textrm{P}} = 2\sum_{i=1}^{32} \, c_{{\rm p},i} \, I^{(n_i)}_{{\rm p},i} \,,
\end{equation}
where the prefactor $2$ is the normalisation stemming from the permutational sum,\footnote{Note that unlike the nonplanar case, there is no color factor contribution.} and the UT integrals are (as in nonplanar case, we only indicate the numerator)
\allowdisplaybreaks[1]
\begin{align}
\I{1}{{\rm p},1}= \ & \ppp{1}{19}{2} \\[0.2em]
 \I{2}{{\rm p},2}= \ & (-\pp{2}{2}-\pp{2}{11}+\pp{2}{19})\pp{2}{19}  \\[0.2em]
 \I{3}{{\rm p},3}= \ & (\pp{3}{1}+\pp{3}{3}+\pp{3}{9}+\pp{3}{10}-\pp{3}{19})\pp{3}{19} \\[0.2em]
 \I{4}{{\rm p},4}= \ & (-\pp{4}{2}-\pp{4}{10}+\pp{4}{19}) \pp{4}{19} \\[0.2em]
 \I{5}{{\rm p},5}= \ & (\pp{5}{2}+\pp{5}{9}-\pp{5}{19})^2-4 \pp{5}{2} \pp{5}{9} \\[0.2em]
 \I{6}{{\rm p},6}= \ & (\pp{6}{3}+\pp{6}{10}-\pp{6}{18}-\pp{6}{19})\pp{6}{19}  
\\
 \I{7}{{\rm p},7}= \ & (-\pp{7}{3}-\pp{7}{9}+\pp{7}{19}) 
(-\pp{7}{3}+\pp{7}{5}+\pp{7}{6}+\pp{7}{17}+\pp{7}{19}) \\[0.2em]
 \I{9}{{\rm p},8}= \ & \pp{9}{13} \pp{9}{19} \\[0.2em]
 \I{10}{{\rm p},9}= \ & (2 \pp{10}{5}+2 \pp{10}{7}+\pp{10}{9}-2 \pp{10}{15}) 
(\pp{10}{2}+\pp{10}{10}-\pp{10}{19}) \\[0.2em]
 \I{12}{{\rm p},10}= \ & 
(-\pp{12}{1}-\pp{12}{2}+\pp{12}{3}+\pp{12}{5}+\pp{12}{7}-\pp{12}{8}+\pp{12}{10}-\pp{12}{13}+\pp{12}{14}+\pp{12}{15}  \nonumber\\ &
-\pp{12}{17}+\pp{12}{18}+
\pp{12}{19}) (\pp{12}{5}+\pp{12}{6}-\pp{12}{11}+\pp{12}{14}+\pp{12}{18}
+\pp{12}{19})   \nonumber\\ &
-\pp{12}{3} \pp{12}{5}-\pp{12}{6} 
\pp{12}{10}-\pp{12}{8} \
\pp{12}{11} \\[0.2em]
 \I{12}{{\rm p},11}= \ &
 (-\pp{12}{1}-\pp{12}{2}+\pp{12}{3}+\pp{12}{4}+\pp{12}{5}-
2 \pp{12}{8}+\pp{12}{9}+\pp{12}{10}-\pp{12}{12}-\pp{12}{13}
 \nonumber\\ &
 +\pp{12}{\
14}-\pp{12}{17}+\pp{12}{18}) (2 \pp{12}{1}+2 
\pp{12}{2}-\pp{12}{5}-\pp{12}{6}+2 
\pp{12}{11}-\pp{12}{14}-\pp{12}{18}
 \nonumber\\ & 
 -2 \pp{12}{19}) 
+\pp{12}{3} \pp{12}{5}+\pp{12}{1} \pp{12}{7}+\pp{12}{6} 
\pp{12}{10}+4 \pp{12}{8} \pp{12}{11}+\pp{12}{2} 
\pp{12}{12} \\[0.2em]
 \I{13}{{\rm p},12}= \ & (\pp{13}{4}+\pp{13}{9}-\pp{13}{14}-\pp{13}{19}) 
(-\pp{13}{3}+\pp{13}{7}+\pp{13}{15}+\pp{13}{19}) \\[0.2em]
 \I{14}{{\rm p},13}= \
&(-\pp{14}{4}+\pp{14}{6}+\pp{14}{7}+\pp{14}{16}+\pp{14}{19}) 
(-\pp{14}{3}+\pp{14}{6}+\pp{14}{17}+\pp{14}{19}) \\[0.2em]
 \I{17}{{\rm p},14}= \
&(\pp{17}{2}-\pp{17}{3}+\pp{17}{4}+\pp{17}{6}-\pp{17}{7}-\pp{17}{9}+\pp{17}{11}-\pp{17}{15}+\pp{17}{17})  \nonumber\\ &
(\pp{17}{2}-\pp{17}{7}-\pp{17}{15}-\pp{17}{19}) \\[0.2em]
 \I{17}{{\rm p},15}= \
&(\pp{17}{3}-\pp{17}{4}+\pp{17}{7}-\pp{17}{11}+\pp{17}{15}) 
(-\pp{17}{2}+\pp{17}{3}-\pp{17}{4}-\pp{17}{5}-\pp{17}{11}
 \nonumber\\ &
 +\pp{17}{12}+
\pp{17}{13}+\pp{17}{15}-\pp{17}{17}+\pp{17}{19}) \\[0.2em]
 \I{19}{{\rm p},16}= \ & 
(\pp{19}{2}-\pp{19}{4}-\pp{19}{11}+\pp{19}{13}-\pp{19}{14}) 
(-\pp{19}{12}+\pp{19}{14}+\pp{19}{19}) + \pp{19}{2} \pp{19}{11}\\[0.2em]
 \I{19}{{\rm p},17}= \ & (2 \pp{19}{2}-2 \pp{19}{6}+2 \
\pp{19}{10}-\pp{19}{11}+2 \pp{19}{13}-2 \pp{19}{14}-2 \pp{19}{16}-2 
\pp{19}{19})
 \nonumber\\ &
(\pp{19}{14}-2 \pp{19}{2})  -\pp{19}{2} \pp{19}{11} \\[0.2em]
 \I{21}{{\rm p},18}= \ & (\pp{21}{4}+\pp{21}{10}-\pp{21}{13}-\pp{21}{19})^2 \\[0.2em]
 \I{25}{{\rm p},19}= \ & (\pp{25}{5}+2 \
\pp{25}{6}+\pp{25}{8}-\pp{25}{13}-\pp{25}{16})^2-4 \pp{25}{6} 
\pp{25}{11} \\[0.2em]
 \I{30}{{\rm p},20}=
\ & (-\pp{30}{3}+\pp{30}{4}-\pp{30}{5}-\pp{30}{13}+\pp{30}{15}) 
(\pp{30}{4}-\pp{30}{5}-\pp{30}{8}+\pp{30}{9}-\pp{30}{13}\nonumber\\
& -\pp{30}{19}) \\[0.2em]
 \I{13}{{\rm p},21}= \ & (\pp{13}{6}-\pp{13}{8}+\pp{13}{17})\pp{13}{5}  \\[0.2em]
 \I{14}{{\rm p},22}= \ & \pp{14}{7} \pp{14}{18} \\[0.2em]
 \I{14}{{\rm p},23}= \ & (\pp{14}{1}-2 (\pp{14}{3}-\pp{14}{4}+3 
\pp{14}{7}+\pp{14}{14}-\pp{14}{15}))\pp{14}{6}  \\[0.2em]
 \I{14}{{\rm p},24}= \ & (-2 \pp{14}{3}+2 \pp{14}{5}+2 \pp{14}{6}-2 
\pp{14}{8}+\pp{14}{11}+2 \pp{14}{17}+2 \pp{14}{18}+2 \pp{14}{19}) \pp{14}{10} \\[0.2em]
 \I{17}{{\rm p},25}= \ & (\pp{17}{5}-\pp{17}{12}) (\pp{17}{2}-2 
\pp{17}{3}+\pp{17}{16}+\pp{17}{17}) \\[0.2em]
 \I{17}{{\rm p},26}= \ & (4 \pp{17}{3}-\pp{17}{4}+2 
\pp{17}{7}+\pp{17}{19}) \pp{17}{8} \\[0.2em]
 \I{17}{{\rm p},27}= \ &(-\pp{17}{2}+2 \pp{17}{3}+2 
\pp{17}{5}+\pp{17}{7}+2 \pp{17}{9}-2 \pp{17}{10}-\pp{17}{11}-2 
\pp{17}{14}
 \nonumber\\ &
 +\pp{17}{15}+2 \pp{17}{18}+\pp{17}{19})\pp{17}{7}  \\[0.2em]
 \I{17}{{\rm p},28}= \ &
(\pp{17}{2}-\pp{17}{3}+\pp{17}{4}-\pp{17}{7}+\pp{17}{11}-\pp{17}{13}-
\pp{17}{15}+\pp{17}{17}-\pp{17}{19})\pp{17}{11}  \\[0.2em]
 \I{19}{{\rm p},29}= \ &\pp{19}{10} \pp{19}{14} \\[0.2em]
 \I{19}{{\rm p},30}= \ &(-2 \pp{19}{2}-2 \pp{19}{3}+\pp{19}{19})\pp{19}{1}  \\[0.2em]
 \I{19}{{\rm p},31}= \ &(3 \pp{19}{2}-4 \pp{19}{3}+4 \pp{19}{7}+6 
\pp{19}{9}-3 \pp{19}{14}+4 \pp{19}{17})\pp{19}{11}  \\[0.2em]
 \I{30}{{\rm p},32}= \ &(-4 \pp{30}{3}+4 \pp{30}{4}-4 
\pp{30}{5}+\pp{30}{11}-4 \pp{30}{13}+4 \pp{30}{15})\pp{30}{9} \, ,
\end{align}
where we use propagator basis $D_i^{(n)}$ given in appendix \ref{app:basis}; note we also define $D_{19}^{(n)} = (p_1+p_2)^2$. 

The full four-loop form factor can be obtained as:
\begin{equation}
\label{eq:FFfull}
F^{(4)} = F^{(4)}_{\textrm{P}} + F^{(4)}_{\textrm{NP}} \,.
\end{equation}

\section{Numerical integration in the nonplanar sector}
\label{sec:numeric}

Although preferably one would want analytic results for the integrals that appear in the four-loop form factor, they appear to be somewhat beyond the current state-of-the-art for computing integrals analytically. Two promising analytic approaches are: (1) the detour via introducing an additional scale and subsequent use of differential equations~\cite{Henn:2016men,Lee:2016ixa,Lee:2017mip}, and (2) the finite integral approach of~\cite{vonManteuffel:2014qoa}. In both approaches, the IBP reduction seems to be the main bottleneck. For instance, the latter would also enable the use of dimensional recurrences \cite{Tarasov:1997kx,Lee:2010cga}, but requires the solution to the IBP problem of so-called four-dotted integrals.

In the present work, we choose a numerical approach. While numerical integration of the four-loop form factor integrals remains quite hard for generic numerators, we make the surprising empirical observation that UT integrals are numerically much easier to integrate than generic numerators of the class under study.
We may offer an intuitive explanation for this. The constant leading transcendentality criterion used to find candidate UT integrals guarantees that these integrals have very mild singularity properties. An algorithm like sector decomposition is bound to be more efficient in cases where internal singularities are simpler. Note however that  sector decomposition algorithm works in Feynman parameter space, whereas constant leading singularity criteria are applied in parametric form like \eqref{eq:parametric-form}. Whatever the precise origin, the relative simplicity of UT integrals is a boon for explicit computation, leading to a remarkable reduction in intermediate expression sizes and integration times. Moreover, the obtained coefficients in the expansion appear to be numerically much smaller than for generic integrals; this is beneficial for reducing potential cancellation errors.

Because of the physical motivation, we will only focus on integration of the integrals in the nonplanar sector of the form factor. We leave integration of the integrals in the colour-planar sector to future work, mostly because all terms of the latter through to ${\cal O}(\epsilon^{-1})$ are already known: The $\epsilon^{-\{8,6,5,4,3\}}$ poles are dictated by contributions from lower loops according to eq.~\eqref{eq:logFF}, and the cusp~\cite{Beisert:2006ez,Bern:2006ew,Henn:2013wfa} and collinear~\cite{Cachazo:2007ad, Dixon:2017nat} anomalous dimensions are already known analytically.

\subsection{Mellin-Barnes representations}

Mellin-Barnes (MB) representations constitute a powerful tool for evaluating Feynman integrals~\cite{Smirnov:1999gc,Tausk:1999vh,Anastasiou:2005cb}. They rely on the fact that one can factorise sums of terms at the cost of introducing line integrals in the complex plane. The basic formula reads
\begin{eqnarray}
\frac{1}{\left( A_1+A_2+\,\ldots\, +A_m\right)^{\alpha}} &=& \MB{c_1}{w_1} \, \cdots \hs{-10} \MB{c_{m-1}}{w_{m-1}} \, A_1^{w_1} \,\cdots\,
A_{m-1}^{w_{m-1}} \, A_m^{-\alpha-w_1-\,\ldots\,-w_{m-1}} \nonumber\\[0.5em]
&&\times \frac{\Gamma(-w_1)\,\cdots\,\Gamma(-w_{m-1})\,\Gamma(\alpha+w_1+\,\ldots\,+w_{m-1})}{\Gamma(\alpha)} \; .
\label{eq:MBbasic}
\end{eqnarray}
The curves are usually straight lines parallel to the imaginary axis whose constant real parts are chosen such as to separate all left from all right poles of $\Gamma$-functions. This is achieved by choosing the real parts of all MB variables $w_i$ together with that of $\epsilon$ such that the arguments of all $\Gamma$-functions have positive real parts. The poles in $\epsilon$ are then extracted by analytical continuation to $\epsilon \to 0$, for which several algorithms exist~\cite{Smirnov:2004ym,Smirnov:2006ry,Anastasiou:2005cb,Czakon:2005rk,Smirnov:2009up}. Subsequently, the terms can be Laurent-expanded about $\epsilon = 0$ and integrated, which proceeds mostly numerically with {\tt{MB.m}}~\cite{Czakon:2005rk}, but also examples of analytical evaluation of MB integrals exist~\cite{Heinrich:2007at,Heinrich:2009be}.

To get from a loop integral to an MB representation, one first introduces Feynman parameters, e.g.\ like
\begin{align}
&\int \! d^Dk_1 \, \frac{1}{k_1^2 \, (k_1+q_1)^2 \, (k_1+q_2)^2 \, (k_1+q_3)^2} = \Gamma(4) \int \! d^Dk_1 \int_0^1 \!dx_1dx_2dx_3dx_4 \, \nonumber\\[0.5em]
&\times \frac{\delta(1-x_1-x_2-x_3-x_4)}{\left[k_1^2+x_1x_2q_1^2+x_1x_3q_2^2+x_1x_4q_3^2+x_2x_3(q_1-q_2)^2+x_2x_4(q_1-q_3)^2+x_3x_4(q_2-q_3)^2\right]^4} \, .
\end{align}
After integration over $k_1$ the remaining terms can be factorised using eq.~(\ref{eq:MBbasic}), and subsequently integrated over the $x_i$ via
\begin{equation}
\displaystyle\int\limits_0^1 dx_1 dx_2 \cdots dx_n \; x_1^{a_1-1} \, x_2^{a_2-1} \, \cdots \, x_n^{a_n-1} \,
\delta(1-\sum_{i=1}^n x_i)  = \frac{\Gamma(a_1)\,\Gamma(a_2) \, \cdots \, \Gamma(a_n)}{\Gamma(a_1+a_2+\ldots+a_n)} \; .\label{eq:fpint}
\end{equation}

The procedure is then repeated until all loop momenta are integrated out. In our case where no kinematic thresholds are present one has to obtain positive definite terms at all stages of the calculation if $q^2$ is space-like (we put $q^2=-1$ for definiteness). Moreover, all terms in the $\epsilon$ expansion are real. For planar topologies this so-called {\emph{loop-by-loop approach}} is always applicable, and we will refer to MB representations coming exclusively from positive definite terms as {\emph{valid}} MB representations. However, valid MB representations for a given loop integral are not unique, even their dimensionality can differ depending on the order the loop momenta are integrated over.

For crossed topologies, the situation is more complicated as one encounters cases in which the loop-by-loop approach yields polynomials in the Feynman parameters $x_i$ which are not positive definite, even in the absence of kinematic thresholds. Consequently, the MB integrals will be highly oscillating and hence their numerical evaluation will be difficult to handle, although steps in this direction have been undertaken~\cite{Blumlein:2014maa,Dubovyk:2015yba,Dubovyk:2016ocz,Dubovyk:2017cqw}.

One way of circumventing this problem in the case of crossed topologies is to not integrate over the loop momenta one by one, but to simultaneously integrate over all loop momenta. This is done by means of the Symanzik graph polynomials $\cal U$ and $\cal F$~\cite{Nakanishi:1971,Itzykson:1980rh,Tarasov:1996br,Heinrich:2008si}, which at $L$ loops are homogeneous of order $L$ and $L+1$, respectively. In the absence of kinematic thresholds they are positive definite and hence automatically lead to valid MB representations. The price to pay is the fact that the number of terms in $\cal U$ and $\cal F$ scales as $L!$. As $L$ increases this therefore quickly leads to MB representations that are too high-dimensional to be integrated in practice. A partial remedy to this problem is to group the lengthy sum of terms $x_i x_j \ldots$ in $\cal U$ and $\cal F$ into a short sum of products $(x_i + \ldots)(x_j+\ldots)\ldots$, see our example below.

To take advantage of both the loop-by-loop and the $\cal F$$\cal U$ approach we apply a hybrid of the two approaches here (see also~\cite{Blumlein:2014maa}): To keep the dimension of the MB representation at a manageable level, we first integrate via the loop-by-loop method over as many (say, $\ell$) loop momenta as possible such as to not generate non-positive definite expressions. Afterwards, we use the $\cal F$$\cal U$ method for the remaining $4-\ell$ loops. This ensures that we deal with positive definite terms at all stages of the calculation and hence automatically obtain valid MB representations, and still keep the dimension of the resulting MB integral moderate since the number of terms in $\cal U$ and $\cal F$ only scales as $(4-\ell)!$ instead of $4!$. To give examples we look at different integral topologies from the nonplanar part of the four-loop form factor.

In topology~$(24)$ (see Fig.~\ref{fig:NPtops}) for instance, we can integrate out the box that is attached to the external $p_2$-line. Contrary to expectations the loop-by-loop approach fails if one tries to integrate out next the box attached to the external $p_1$-line or any other loop. Consequently, the remaining three loop momenta have to be treated by means of the $\cal F$$\cal U$ approach. The obtained MB representation is $21$--fold and hence at the edge of what is doable in practice. We did some checks through to ${\cal O}(\epsilon^{-4})$, where at most $7$-fold integrals appear in the numerical evaluation with {\tt{MB.m}}~\cite{Czakon:2005rk}. Topology~$(23)$ with numerator $\left[(\ell_3-p_1)^2\right]^2$ is even worse since we didn't find a single loop that can be integrated over before one is enforced to switch to the $\cal F$$\cal U$ method. Consequently, the MB method was not applied to this topology. In topology~$(25)$, on the other hand, one can integrate over the two boxes attached to the external $p_1$ and $p_2$-lines, respectively, and switch to the $\cal F$$\cal U$ method afterwards. Still, the obtained MB representation is $20$--fold. We use it for some checks through to ${\cal O}(\epsilon^{-5})$, where at most $5$-fold integrals appear in the numerical evaluation.

One example where the hybrid method works particularly well is topology~$(30)$ with numerator $(\ell_3-\ell_4-p_2)^2 \, [(p_1-\ell_4)^2+(\ell_3-\ell_4)^2-(\ell_3-p_1)^2]$, i.e.~$I^{(30)}_{11}$. Let us therefore give more details on this case. After shifting the loop momenta according to
\begin{align}
\ell_3 &= p_1 - k_4 , & \qquad \ell_5 & = k_3 - k_1 , \nonumber\\
\ell_4 &= p_1 - k_3 , & \qquad \ell_6 & = k_2 ,
\end{align}
the numerator becomes $(k_3-k_4-p_2)^2 \, [k_3^2+(k_3-k_4)^2-k_4^2]$. We now integrate out the two boxes parameterised by the loop momenta $k_2$ and $k_1$. Using eqs.\ (\ref{eq:MBbasic})~--~(\ref{eq:fpint}) this introduces eight MB parameters and we are left with the propagators
\begin{align}
\{(k_4-p_1)^2 , (k_4+p_2)^2 , (k_3-k_4)^2 ,  k_3^2 , (k_3-p_1)^2 , (k_3+p_2)^2 , (k_3-k_4-p_2)^2 \}
\label{eq:propsFU}
\end{align}
that are raised to various powers which can also depend on the MB variables. At this stage it becomes obvious that all terms in the numerator except $k_4^2$ can be treated as inverse propagators, giving rise to shifted propagator powers. While in principle also $k_4^2$ could be treated in this way, it would introduce an additional propagator and hence longer $\cal F$ and $\cal U$ polynomials. We therefore use the formulas in section~$3.2.4$ of~\cite{Borowka:2015mxa} (which are based on ideas in~\cite{Smirnov:2006ry,Smirnov:2008py}) for explicit numerator factors in case of $k_4^2$.

The $\cal F$ and $\cal U$ graph polynomials for this two-loop topology can now be written down, and a crucial step consists of writing the expanded sum of terms as a short sum of products which still happens to be positive definite. One obtains
\begin{align}
\allowdisplaybreaks
{\cal U} \; = & \; x_1 x_3+x_2 x_3+x_4 x_3+x_5 x_3+x_6 x_3+x_1 x_4+x_2 x_4+x_1 x_5 \nonumber \\[0.3em]
           & +x_2 x_5 +x_1 x_6+x_2 x_6+x_1 x_7+x_2 x_7+x_4 x_7+x_5 x_7+x_6 x_7 \nonumber \\[0.3em]
	   = & \; \left(x_1+x_2\right) \left(x_4+x_5+x_6\right)+\left(x_3+x_7\right) \left(x_4+x_5+x_6\right)+\left(x_1+x_2\right) \left(x_3+x_7\right) \, , \nonumber \\[0.3em]
{\cal F} \; = & \; x_1 x_2 x_3+x_2 x_5 x_3+x_1 x_6 x_3+x_5 x_6 x_3+x_1 x_2 x_4+x_1 x_2 x_5 \nonumber \\[0.3em]
              & +x_1 x_2 x_6+x_1 x_5 x_6+x_2 x_5 x_6+x_1 x_2 x_7+x_1 x_4 x_7+2 x_1 x_6 x_7+x_5 x_6 x_7 \nonumber \\[0.3em]
	    = & \; x_2 x_5 \left(x_3+x_6\right)+x_1 x_2 \left(x_4+x_5+x_6\right)+x_1 \left(x_4+x_6\right) x_7 \nonumber \\[0.3em]
	      & +\left(x_1 x_2+x_5 x_6\right) \left(x_3+x_7\right)+x_1 x_6 \left(x_3+x_5+x_7\right) \, ,
\label{eq:FUtop30}
\end{align}
and hence $\cal U$ and $\cal F$ are sums of three and five terms only, respectively. Moreover, they do have various factors such as $(x_4+x_5+x_6)$ and $(x_3+x_7)$ in common. The factorisation of the terms in $\cal F$ and $\cal U$ via eq.~(\ref{eq:MBbasic}) now proceeds in several steps. The final integration over the $x_i$ is performed by means of eq.~(\ref{eq:fpint}) and requires the introduction of an additional regulator $\delta$ in order to avoid a $\Gamma(0)$ in the denominator. We choose to add $\delta$ to the power of $x_4$.

After application of Barnes' lemmas the resulting MB representation is $14$-fold. The subsequent analytic continuation $\delta \to 0$ is done with {\tt MB.m}~\cite{Czakon:2005rk} and the dimension is reduced to~$13$. This integrand is attached to the arXiv submission of the present article.

The package {\tt MB.m} is also used for most of the remaining steps: Analytic continuation to $\epsilon = 0$, expansion in $\epsilon$, application of Barnes' lemmas, and numerical integration. After these steps, at most six-fold MB integrands appear through to ${\cal O}(\epsilon^{-2})$. At ${\cal O}(\epsilon^{-1})$ the maximum dimension of the integrand is seven. We use the algorithms CUHRE and VEGAS from the CUBA library with up to $2.1$~billion sampling points. The result is given in appendix~\ref{sec:utints}.

\subsection{Sector decomposition}

Sector decomposition \cite{Binoth:2000ps,Heinrich:2008si} regularises the $\epsilon$-expansion of Feynman integrals by performing a blow-up at singularities of the Feynman parameter representation of a given integral. Sector decomposition has been implemented in several public codes, e.g.\ FIESTA  \cite{Smirnov:2008py, Smirnov:2009pb, Smirnov:2013eza, Smirnov:2015mct} and SecDec \cite{Carter:2010hi, Borowka:2012yc, Borowka:2015mxa}. For our production runs, we have used the FIESTA code, with cross-checks in simpler cases from SecDec. After resolving the singularities, a list of remaining integrals is obtained. These could be integrated analytically in principle, but most often these are integrated numerically. The numerical integration is performed using mainly the VEGAS algorithm \cite{Lepage:1980dq} as implemented in the CUBA library \cite{Hahn:2004fe}, with some cross-checks using CUHRE and DIVONNE from the same library. For our production runs, we have typically used several 100 million sampling points per integral in the Monte Carlo-based numerical integration algorithms. This requires large computing resources. 
 
In the course of computation several tricks were used to speed up computation and to control the arising errors. The sector decomposition programs involve choices of how to regularise the integrals, which are encapsulated in different strategies for resolving the singularities. This is a feature of which the problem at hand benefits a lot since the occurring integrals are complicated multivariate expressions. Whenever it finishes, FIESTA's ``strategy X'' typically leads to smallest sector counts which we will take to be a proxy for the ease of integration. In cases where this strategy fails for one or more sectors, one can split the computation into those sectors treated with strategy X and a remainder tackled with ``strategy S''. This can be done using the option ``SectorCoefficients'' in FIESTA. A further trick to use is that of graph symmetries. These can be used to gather exponents of several sectors into a single one, with a numerical pre-factor counting the number of sectors related to the base sector. For choosing the base sector, one first runs FIESTA on all sectors, selecting representatives which have the smallest sector count as a proxy for simplicity. 

Note it is very important to verify that the integral in question has the explicit graph symmetry used; otherwise a wrong result may be the consequence. Here the UT properties of the integrals offer some protection: if an error with respect to graph symmetries is made during computation, the obtained final result is typically not UT and this manifests itself for instance by a non-vanishing $\epsilon^{-7}$ coefficient. Moreover, in these cases the numerical value of the coefficients tends to grow very fast with increasing orders of $\epsilon$. In addition, if a graph symmetry is misused one cannot rationalise the coefficients of the $\epsilon$ expansion as described below. 

\begin{table*}[t]
\caption{Nonplanar form factor result and errors. The prefactor ${48 /N_c^2}$ in \eqref{eq:FF_NP} is not included.}
\label{tab:error}
\centering
\vspace*{5pt}

\resizebox{0.8\textwidth}{!}{
\begin{tabular}{l | c| c| c | c } 
\hline\hline
$\epsilon$ order & \, $-$8 \, & \, $-$7 \,& \, $-$6 \,& \, $-$5 \,    \cr \hline 
result           & $-3.8\times 10^{-8}$ & $+4.4\times 10^{-9}$ & $-1.2\times 10^{-6}$ & $-1.2\times 10^{-5}$ \cr \hline 
uncertainty      & -- & $\pm 5.7\times 10^{-7}$ & $\pm 1.0\times 10^{-5}$ & $\pm 1.2\times 10^{-4}$ \cr \hline \hline
\end{tabular}}

\vspace*{8pt}

\resizebox{0.62\textwidth}{!}{
\begin{tabular}{l | c| c| c | c} 
\hline\hline
$\epsilon$ order & \, $-$4 \,  & \, $-$3 \, & \, $-$2 \, & \, $-$1 \,    \cr \hline 
result   	 &  $+3.5\times 10^{-6}$ & $+$~0.0007 & $+$1.60 & \,$-$17.98 \,   \cr \hline 
uncertainty    	 &  $\pm 1.5\times 10^{-3}$  & $\pm$ 0.0186  & $\pm$0.19 & $\pm$ 3.25 \cr \hline \hline
\end{tabular}}

\end{table*}

\subsection{Nonplanar cusp and collinear anomalous dimensions}
We gather the numerical results for all integrals needed for the nonplanar part of the Sudakov form factor in appendix~\ref{sec:utints}. When combined to give the Sudakov form factor, the results are gathered in table~\ref{tab:error}. Errors are added in quadrature, see below for the rationale behind this. Due to the high precision of the computation at order $\epsilon^{-8}$, there is no sensible reported error in FIESTA. Note that in table~\ref{tab:error} the prefactor ${48 /N_c^2}$ in \eqref{eq:FF_NP} is not included.

As mentioned above, physics dictates that the coefficients of orders $\epsilon^{\{-8,-7,-6,-5,-4,-3\}}$ vanish in the final result, which is numerically indeed the case and provides a strong consistency check of our computation. The coefficients of order $\epsilon^{-7}$ must even vanish in each of the $23$ UT integrals separately. The orders $\epsilon^{\{-8,-6,-5,-4,-3\}}$ are in most cases non-zero in individual integrals but cancel in the final result. As described below, the precision of the orders $\epsilon^{\{-8,-6,-5,-4\}}$ is good enough to translate the reported numbers into \emph{small} rational multiples of $\{1,\zeta_2,\zeta_3,\zeta_4\}$. After doing so, these orders also vanish analytically in the final result of the nonplanar form factor.

As can be seen from table~\ref{tab:error}, the first non-zero term is at order $\epsilon^{-2}$. The result $1.60 \pm 0.19$ has a statistical significance to deviate from zero of~$8.4 \sigma$. Adding individual uncertainties linearly to account for potential systematic effects would yield $1.60 \pm 0.58$; still significantly non-zero.\footnote{Note that these numbers are slightly improved compared to those in~\cite{Boels:2017skl}.} We will argue below that there is no evidence for systematically underestimated error bars in our calculation.

Translating the result of the order $\epsilon^{-2}$ of the nonplanar form factor into a result for the sought-after nonplanar four-loop CAD yields for gauge group $SU(N_c)$
\begin{equation}
\label{eq:CAD-result}
\gamma_{\textrm{cusp, NP}}^{(4)}  = -3072\times( 1.60 \pm 0.19 ) \frac{1}{N_c^2} \,,
\end{equation}
where the prefactor $3072 = 2 \times 24 \times 64 $ is the normalisation stemming from the permutational sum, the colour factor~\cite{Boels:2012ew}, and the denominator of~(\ref{eq:centralrelation}), respectively. Compared to the planar result $\gamma_{\rm cusp, P}^{(4)} = - 1752 \zeta_6 - 64\zeta_3^2\sim-1875$, we observe that the nonplanar CAD has the same sign. If we use $N_c=3$, its value becomes $\gamma_{\textrm{cusp, NP}}^{(4)} \sim - 546 \pm 65$, i.e.\ the planar contribution is a factor of 3~--~4 larger.

The result at order $\epsilon^{-1}$ is also given in table~\ref{tab:error}. This contains the nonplanar four-loop collinear anomalous dimension:
\begin{equation}
\label{eq:CollAD-result}
{\cal G}_{\textrm{coll, NP}}^{(4)}  = -384\times( -17.98 \pm 3.25 ) \frac{1}{N_c^2} \,,
\end{equation}
where the prefactor $384 = 2 \times 24 \times 8 $ has the similar origin as $\gamma_{\textrm{cusp, NP}}^{(4)}$ above.
Interestingly, compared to the four-loop planar collinear AD result, ${\cal G}_{\textrm{coll, P}}^{(4)} \sim -1240$ \cite{Cachazo:2007ad, Dixon:2017nat}, we observe that the nonplanar central value result $+ (6904 \pm 1248)/N_c^2$ indicates the sign is different; it is also different from the sign of the nonplanar cusp AD above. This is a new feature comparing to all known planar results in which collinear AD always has same sign as cusp AD.\footnote{One should also keep in mind that unlike cusp AD, collinear AD is scheme dependent, thus the sign may change in different schemes.}
Note that our result is in tension with a vanishing result at the ~$5.5\sigma$ level. The largest contribution to the error budget within the integrals at this order comes from $I_8^{(27)}$, which contributes $\sim 1.86$, followed by four integrals which contribute between~$0.95$ and~$1$ each, whereas all others are below~$0.75$. We mention that the linearly summed error is obtained as $-17.98 \pm 11.89$.

To improve the reported uncertainties significantly within our numerical approach would come at a high price, both with respect to computing time and power, since the resources required for pushing this computation through sector decomposition are fairly large. It would certainly be interesting though to confirm the sign of the collinear AD.

\subsubsection{Rationalisation}\label{sec:rationalisation}

Since the used integrals pass all applied UT checks, their $\epsilon$-expansion is expected to be UT. Assuming that MZVs are sufficient and no genuine Euler sums occur, for the orders $\epsilon^{\{-8,-6,-5,-4\}}$ it is expected that the numerical coefficients can be written as a rational number times $\{1,\zeta_2,\zeta_3,\zeta_4\}$. Hence, by dividing the numerical result by the appropriate MZV constant, a numerical result is obtained which should be expressible as a rational number. For the case at hand, we typically have at least five to six digits available and the found integers have on the order of three digits in numerator and denominator. This indicates that the obtained rational numbers are reasonable, which gets supported by the fact that their contribution in the final result of the nonplanar form factor cancels exactly. In appendix~\ref{sec:utints} the results of the rationalisation are listed.

For $\epsilon^{\{-3,-2\}}$ the UT property still holds, but at these orders there are two MZVs of transcendentality $5$ and $6$ respectively. For weight $5$ these could for instance be taken to be $\zeta_2 \zeta_3$ and $\zeta_5$, and one can attempt a solution with the PSLQ algorithm~\cite{Ferguson:1999:API:307090.307114}, for instance through {\tt Mathematica}'s command {\tt FindIntegerNullVector}. The appropriate integer relation then contains three unknowns: one for the numerical result, and {\emph{two}} for the MZVs. For integer coefficients to be reliably isolated one needs much more digits in these cases, certainly more than 10. Since we have typically only four to five digits available at these orders, the PSLQ algorithm is currently not feasible. Moreover, many of our numerical results were obtained using sector decomposition where the price of integration roughly scales quadratically with increasing precision. This makes PSLQ unfeasible for the coefficients at orders $\epsilon^{\{-3,-2,-1\}}$ within the numerical setup employed here. It would be highly interesting to obtain high precision numerics at these orders, or even better of course analytic results that do not rely on PSLQ.

\subsubsection{Error analysis}\label{sec:erroranalysis}

Since numerical integration methods are used, a thorough discussion of the errors in these integrals is called for. Both for MB as well as for sector decomposition methods an error is reported. As is well-known, if an efficient MB representation can be found, the error in its integration is in general small, especially compared to sector decomposition. For the integrals at hand typically a difference in precision of three to five digits arises. Hence, the discussion here will focus on sector decomposition.

FIESTA employs the CUBA \cite{Hahn:2004fe} integration library. Although we have cross-checked some simple integrals as well as leading expansion coefficients of more complicated ones, most of the coefficients needed for the cusp anomalous dimension at order $\epsilon^{-2}$ were obtained using exclusively the VEGAS \cite{Lepage:1980dq} algorithm. VEGAS employs an adaptive sampling algorithm. It should be noted that the integrals under study do not have any physical singularities, and do not have to be analytically continued, two common sources of error. For sufficiently many evaluation points, the VEGAS error is of Gaussian type. To check that this regime is reached, one evaluates the integrals for several evaluation points settings. In the Gaussian regime, the error scales as $1/\sqrt{\textrm{eval points}}$. For all integrals in the set integrated here, this was reached very quickly. In rare cases involving much more complicated integrals, it has been reported in \cite{Marquard:2016dcn} that the error in FIESTA can be underestimated. In those cases the central value of certain coefficients changed outside the reported error with increasing evaluation points. We have checked for this as well, and have never observed variations outside of reported error upon increasing the number of evaluation points for the integrals under study. Several simpler integrals have been computed using SecDec with the DIVONNE and CUHRE algorithms as a further crosscheck. More cross-checks for integral $I_1^{(21)}$ and $I_{16}^{(30)}$ follow from available MB results, as well as an exact result for integral $I_1^{(21)}$. 

\begin{figure}
  \subfigure[]{
    \label{fig:subfig:a} 
    \includegraphics[width=7.2cm]{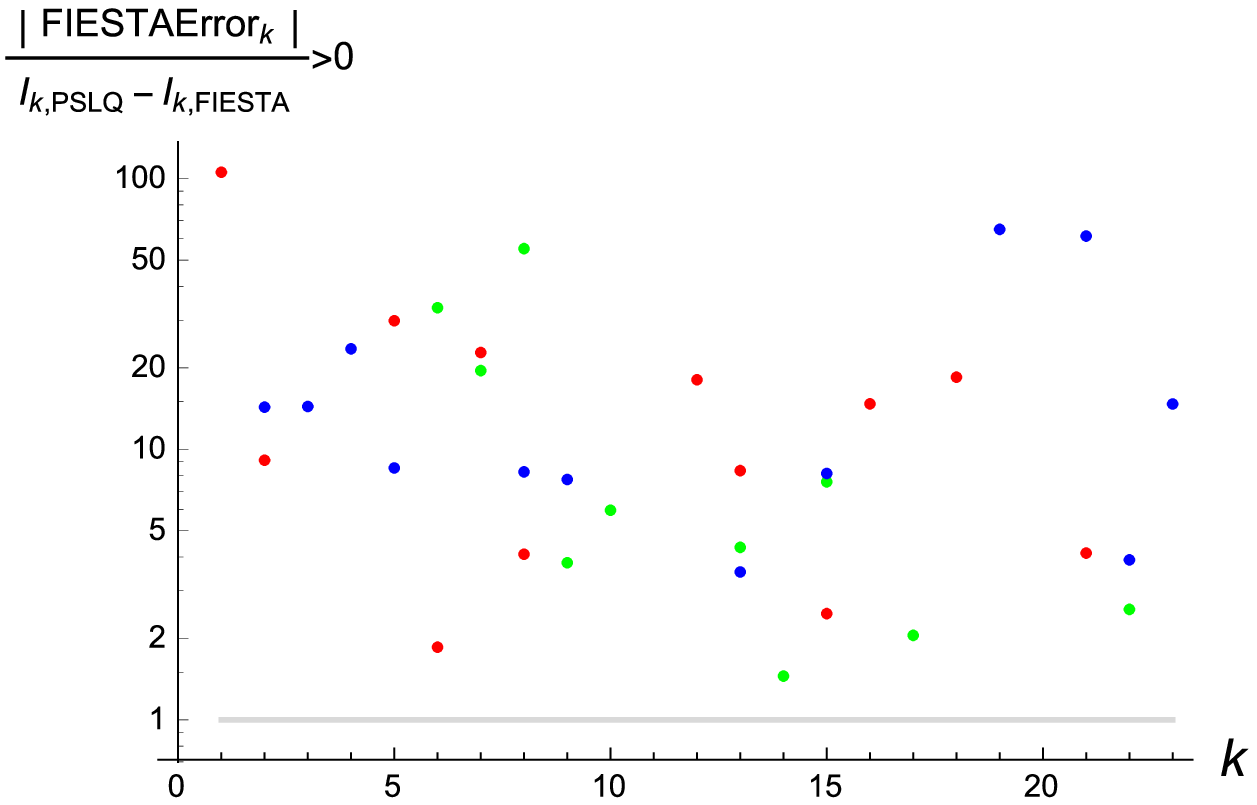}}
  \subfigure[]{
    \label{fig:subfig:b} 
    \includegraphics[width=8.2cm]{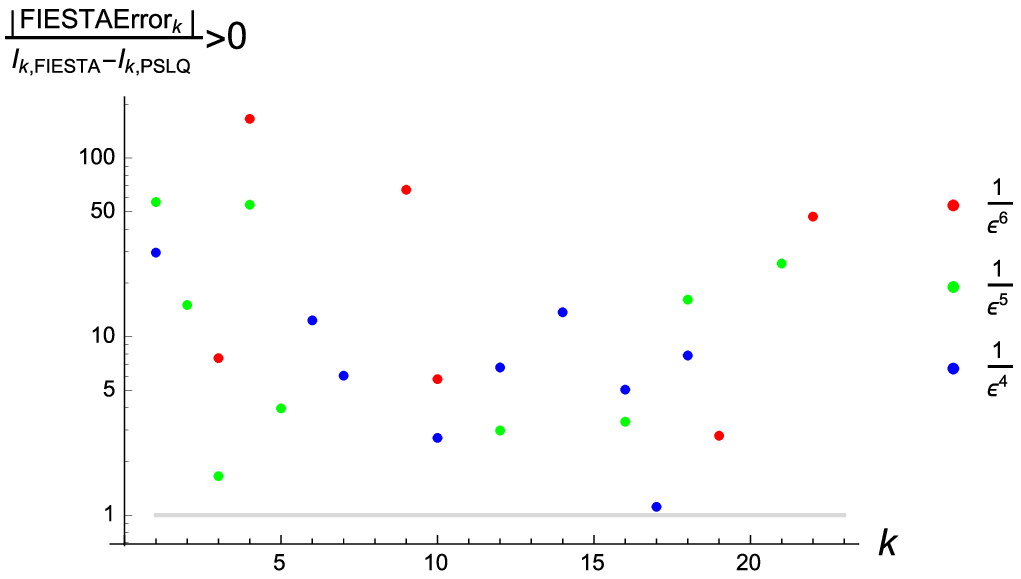}}
  \caption{Scatterplot of the relative error of FIESTA results compared to PSLQ results for $\epsilon^{\{-6,-5,-4\}}$ orders. (a) Plot of cases ${\textrm{FIESTA error} \over I_{\rm PSLQ} - I_{\rm FIESTA}} > 0$. (b)  Plot of cases ${\textrm{FIESTA error} \over I_{\rm FIESTA} - I_{\rm PSLQ}} > 0$. A logarithmic scale is used for the vertical axis, and all ratios larger than $200$ are not shown in the figures. We can see that all ratios are larger than unity, which suggests  that the FIESTA errors are conservative estimates. Besides, we find that the deviation of FIESTA results from PSLQ results are both positive and negative, which indicates that there is no source of systematic errors.}
  \label{fig:PSLQcheck} 
\end{figure}

For the leading coefficients of the individual integrals an additional cross-check is enabled by their UT properties: having obtained a product of a rational number times a zeta value for the leading coefficients from the expansion (see section~\ref{sec:rationalisation}), one can use this to obtain an estimate of the true precision.
For this, we compute the ratio between FIESTA errors and the assumed 'true' errors obtained by comparing to the PSLQ result at order $\epsilon^{\{-6,-5,-4\}}$, namely,
\begin{equation}
{\textrm{FIESTA error}_k \over I_{k,{\rm PSLQ}} - I_{k,{\rm FIESTA}}} \,,
\end{equation}
where $k$ labels the 23 integrals in section~\ref{sec:NPUT}.
The results are plotted in figure~\ref{fig:PSLQcheck}. Two panels are provided for positive and negative deviations separately.
Note that for all $23$ integrals, all absolute ratios are larger than one, corresponding to reported FIESTA errors larger than the discrepancy between PSLQ result and numerical integration.
Moreover, by comparing figure~\ref{fig:PSLQcheck}(a) and figure~\ref{fig:PSLQcheck}(b), it is clear there is no definite sign of the deviation: positive and negative deviations are about as likely. If this had been different, this might have indicated a systematic error.

Finally, physics provides a strong cross-check of the numerics. The leading coefficient of the nonplanar form factor should be of order $\epsilon^{-2}$, while individual integrals generically contribute from order $\epsilon^{-8}$. Hence, in the sum there should be numerical cancellations between the integrals to give zero within error bars for the first six orders of expansion, down to $\epsilon^{-3}$.  With the errors added in quadrature and the result for the sum of the central value, one can compare to the exact answer, $0$, for these coefficients. These results are contained in table \ref{tab:error} and clearly indicate that reported errors are not underestimated, giving further support for our error analysis.

In total, the above analysis shows that the errors reported by FIESTA are stable and in general conservatively estimate the errors for the form factor integrals in the present study. This strongly indicates that the final error for CAD is not underestimated either, and hence there is no need to manually inflate the reported uncertainty. Conservatively, we will interpret the FIESTA reported error as the standard deviation of a Gaussian error. For a true single standard deviation in a Gaussian error, one would expect deviations from the true result to exceed the standard deviation of the Gaussian distribution roughly $32\%$ of the time, while here this never occurs. As a consequence of the error interpretation, the obtained errors are added in quadrature. For reference, also the result of adding errors linearly is provided, which is recommended in cases which involve a small systematic error. However, we emphasise that there is no sign of systematic errors in the case at hand.

\section{Discussion and conclusion}\label{sec:conclusion}

In this article a set of tools and techniques have been discussed for the integration of four-loop form factor integrals, especially focussed on the nonplanar sector of the Sudakov form factor in maximally supersymmetric Yang-Mills theory. This sector contains among others information on the nonplanar correction to the cusp anomalous dimension. Four loops is the first time a nonplanar correction enters into the form factor as well as into the cusp and collinear anomalous dimensions. Although conjectures existed that the CAD vanished generically in gauge theories, our results, first announced in \cite{Boels:2017skl}, show this is not the case. In this article we also present the first numerical result for the nonplanar collinear anomalous dimension. The numerics of especially the latter result leave quite some room for improvement. Even more interesting would be to obtain an analytic result. Besides settling conjectures, of much wider interest is how the results reported in this article were obtained: the tools and techniques are certainly applicable to a wider context than just this particular computation in this particular theory. 

Inspired by similar computations in the literature~\cite{Arkani-Hamed:2014via,Bern:2014kca,Bern:2015ple,Henn:2016men}, an algorithm was presented to find complete sets of uniformly transcendental integrals in a given set of topologies. The algorithm is based on the conjecture that these integrals always have constant leading singularities. Importantly, the algorithm stabilises to a result in finite time in our current {\tt Mathematica} implementation. A surprising amount of uniformly transcendental integrals were found for each integral topology for the problem at hand. With some combination techniques, a set of integrals was obtained to express the maximally supersymmetric form factor in. However, the number of UT integrals involved in this physical problem is much smaller than the total number of UT integrals in each topology. This points towards applications of these integrals beyond maximal supersymmetric Yang-Mills. Intriguingly, the numbers obtained are comparable to the total number of IBP master integrals. It would be very interesting to explore this further, but this will have to involve IBP-reducing the pure, non-supersymmetric Yang-Mills form factor, which is beyond currently (publicly) available technology.

Having obtained a suitable basis of UT master integrals to express the form factor in, the next step is the integration of these integrals. A pleasant surprise is the observation that even though many integration techniques such as sector decomposition  spoil UT properties in intermediate steps, the UT integrals appear to be much easier to integrate than generic integrals in the form factor class. Within sector decomposition, this manifests itself in term counts which are an order of magnitude better. This in turn leads to much more compact expressions in the integration steps which lead to much improved performance in both speed and accuracy. Intuitively, this corresponds well to the notion that UT integrals are inherently simple. More mathematically, the absence of higher order singularities in the integrand in parametric form (as discussed in section 3.1) translates very likely to less singular integrands in Feynman parameter form. This in turn should then explain the observed much improved behaviour of sector decomposition methods. It would be interesting to explore this further, especially a criterion which would allow one to decide if an integral is UT in Feynman parameter form would be highly desired. Since there are considerably fewer integrations in Feynman parameter form than in parametric form, this is potentially even much more powerful. 

Special attention is paid to the numerical integration of the form factor integrals in the nonplanar sector. Apart from the central value, the error analysis in numerical applications is important. Here the UT property of the integrals informs the error analysis. The integration of leading coefficients allows one to check the error analysis by using the PSLQ algorithm to find the exact value of the integrals. This combination of number theory and numerical integration shows that the errors reported by FIESTA are in general very conservative estimates. Added to knowledge of a single exact integrals and several results obtained using Mellin-Barnes integrals, this gives comprehensive evidence for our error analysis for the computation of the nonplanar cusp and collinear anomalous dimensions at four loops. 

\acknowledgments 
It is a pleasure to thank Sven-Olaf Moch, Andreas von Manteuffel and Robert Schabinger for discussions. 
We also thank Volodya Smirnov for bringing to our attention a typo in \eqref{eq:2212lines} in the previous version. 
This work was supported by the German Science Foundation (DFG) within the Collaborative Research Center 676 ``Particles, Strings and the Early Universe''. GY is supported in part by the Chinese Academy of Sciences (CAS) Hundred-Talent Program, by the Key Research Program of Frontier Sciences of CAS, and by Project 11647601 supported by National Natural Science Foundation of China.

\appendix

\section{UT integrals}
\label{sec:utints}

\subsection{UT integrals with 12 lines}
\label{sec:utints12}

For the UT integrals we use the parametrizaton in terms of loop momenta from~\cite{Boels:2015yna} and the normalisation used by FIESTA, i.e.\ we work in $D=4-2\ep$-dimensional Minkowskian space-time and our integration measure is $e^{\ep \gamma_E} \, d^D\ell/(i\pi^{D/2})$ per loop. Moreover, we set $(p_1+p_2)^2 = -1$ and suppress the fact that the $\epsilon$-expansion continues in all equations. Below we give our numerical results as well as the PSLQ up to $\epsilon^{-4}$ order.


\subsection*{Topology~21}
\label{sec:top2112}

\begin{align}
I_{1}^{(21)}  =& \begin{tabular}{c}{\includegraphics[height=1.6cm]{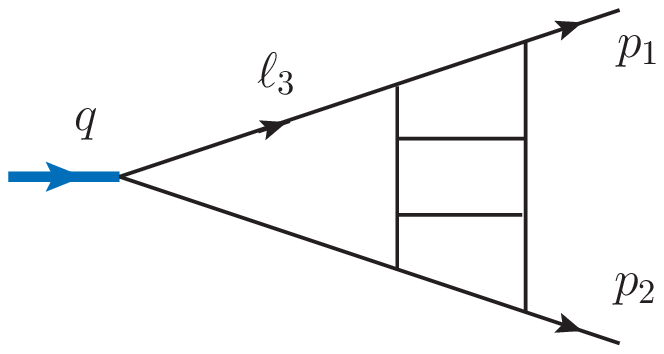}}\end{tabular}   \hskip -.5cm
\times \, [(\ell_3-p_1)^2]^2 
\nonumber \\[1.0em]
                                                  &=\frac{1}{576 \ep^8}+\frac{\zeta_2}{36 \ep^6}+\frac{151 \zeta_3}{864 \ep^5}+\frac{865 \zeta_4}{576\ep^4}+\frac{\frac{505}{216}\zeta_2\zeta_3+\frac{5503}{1440}\zeta_5}{\ep^3}+\frac{\frac{44219}{1152}\zeta_6+\frac{9895}{2592}\zeta_3^2}{\ep^2}
\nonumber \\[1.0em]& +\frac{\frac{89593}{864}\zeta_3\zeta_4+\frac{3419}{45}\zeta_2\zeta_5-\frac{169789}{4032}\zeta_7}{\ep}\, .\label{eq:2112lines}
\end{align}
The integral $I_{1}^{(21)}$ is known analytically from~\cite{Henn:2016men}. Our numerical results obtained by MB and FIESTA agree with the analytical one well within error bars.
\begin{align}
I_{1,{\rm MB}}^{(21)}  = 
& \frac{0.001736111111111111}{\ep^8} +\frac{0.04569261296800628(1)}{\ep^6}+\frac{0.2100817041401606(1)}{\ep^5}\nonumber\\
&  +\frac{1.6253638839586(7)}{\ep^4}+\frac{8.5855125581(10)}{\ep^3}+\frac{44.566338023(40)}{\ep^2} \, , \\[1.0em]
I_{1,{\rm FIESTA}}^{(21)}  =
& \frac{0.00173611}{\ep^8} -\frac{0.0000000004(837)}{\ep^7}+\frac{0.0456926(14)}{\ep^6}+\frac{0.210082(17)}{\ep^5} \nonumber\\
&  +\frac{1.62537(18)}{\ep^4}+\frac{8.5853(19)}{\ep^3}+\frac{44.564(20)}{\ep^2} \, .
\end{align}


\subsection*{Topology~22}
\label{sec:top2212}

\begin{align}
I_{2}^{(22)}  = & \begin{tabular}{c}{\includegraphics[height=1.6cm]{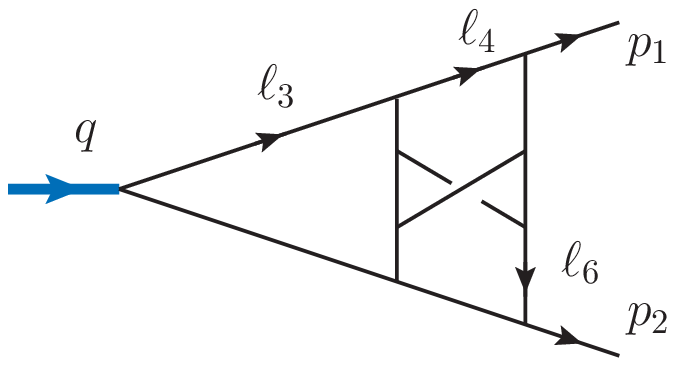}}\end{tabular}   \hskip -.5cm
\times \, \{-(\ell_3-p_1)^2 \, [\ell_4^2+\ell_6^2-\ell_3^2+(\ell_3-\ell_4+p_1)^2+(\ell_3-\ell_6-p_1)^2]\}  \nonumber \\[1.0em]
                 = & \ \frac{0.00520833}{\ep^8}-\frac{0.000000003(130)}{\ep^7}-\frac{0.4340801(26)}{\ep^6}-\frac{2.291419(35)}{\ep^5} -\frac{9.56243(42)}{\ep^4} \nonumber\\
& \ -\frac{51.4505(51)}{\ep^3}-\frac{333.021(67)}{\ep^2}-\frac{1705.78\pm 1.46}{\ep} \, ,\label{eq:2212lines}
\end{align}
\begin{align}
I_{2,{\rm PSLQ}}^{(22)}  =
\frac{1}{192\ep^8}  - \frac{19\zeta_2}{72\ep^6}  - \frac{61\zeta_3}{32\ep^5} - \frac{5089\zeta_4}{576\ep^4} +{\cal O}(\ep^{-3}) \,.
\end{align}


\subsection*{Topology~23}
\label{sec:top2312}

\begin{align}
I_{3}^{(23)}  =& \begin{tabular}{c}{\includegraphics[height=1.6cm]{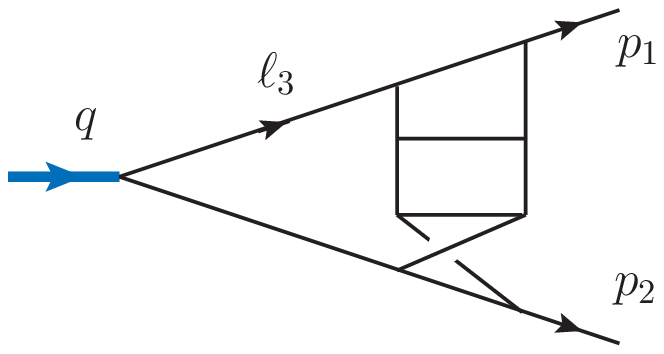}}\end{tabular}   \hskip -.5cm
\times  \, [(\ell_3-p_1)^2]^2  \nonumber \\[1.0em]
                                                 = & \ \frac{0.00694444}{\ep^8}-\frac{0.000000001(45)}{\ep^7}-\frac{0.45692600(98)}{\ep^6}-\frac{2.231590(11)}{\ep^5} +\frac{3.21314(10)}{\ep^4} \nonumber\\
& \ +\frac{73.5027(9)}{\ep^3}+\frac{351.351(8)}{\ep^2} +\frac{664.498(633)}{\ep} \, ,\label{eq:2312lines}
\end{align}
\begin{align}
I_{3,{\rm PSLQ}}^{(23)}  =
& \frac{1}{144\ep^8}  - \frac{5\zeta_2}{18\ep^6}  - \frac{401\zeta_3}{216\ep^5} + \frac{95\zeta_4}{32\ep^4} +{\cal O}(\ep^{-3}) \,.
\end{align}


\subsection*{Topology~24}
\label{sec:top2412}

\begin{align}
I_{4}^{(24)}  =& \begin{tabular}{c}{\includegraphics[height=1.6cm]{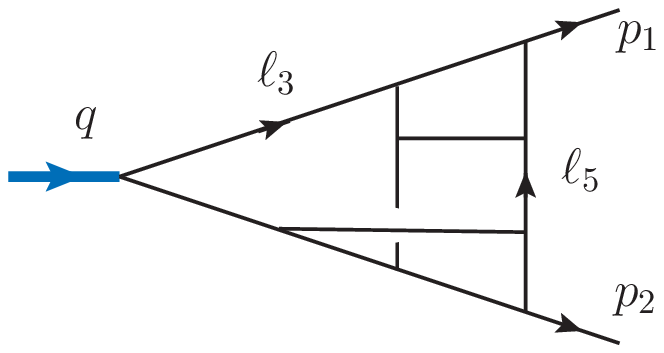}}\end{tabular}   \hskip -.5cm
\times  \, (\ell_3-p_1)^2 \, [(q-\ell_3-\ell_5)^2 + (\ell_5+p_2)^2 ] \nonumber \\[1.0em]
                = & \ -\frac{0.00868056}{\ep^8}+\frac{0.00000002(34)}{\ep^7}+\frac{0.7425050(65)}{\ep^6}+\frac{2.288640(86)}{\ep^5} -\frac{7.37337(101)}{\ep^4} \nonumber\\
& \ -\frac{78.1528(116)}{\ep^3}-\frac{220.386(91)}{\ep^2}+\frac{176.718(990)}{\ep} \, , \label{eq:2412lines}
\end{align}
\begin{align}
I_{4,{\rm PSLQ}}^{(24)}  
& = -\frac{5}{576\ep^8}  + \frac{65\zeta_2}{144\ep^6}  + \frac{1645\zeta_3}{864\ep^5} - \frac{109\zeta_4}{16\ep^4} +{\cal O}(\ep^{-3})  \,.
\end{align}


\subsection*{Topology~25}
\label{sec:top2512}

\begin{align}
I_{5}^{(25)}  =& \begin{tabular}{c}{\includegraphics[height=1.6cm]{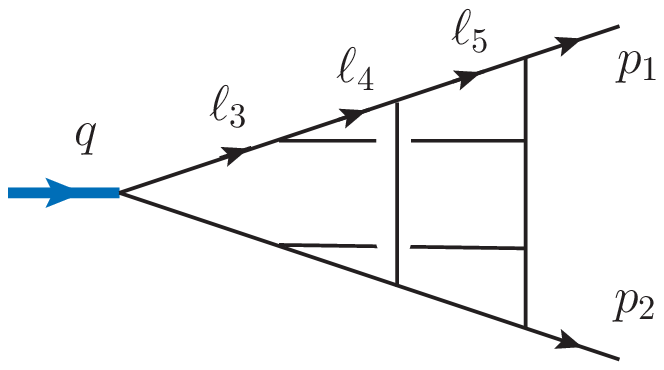}}\end{tabular}   \hskip -.5cm
\times  \, \Big\{\left[(p_1 - \ell_5)^2+2 (\ell_4-\ell_5)^2+(\ell_3-\ell_4)^2-(\ell_3-\ell_5)^2-(p_1-\ell_4)^2 \right]^2 \nonumber \\[1.0em]
                &  \hspace*{25pt} -4\, (\ell_4-\ell_5)^2 \, (p_1-\ell_3+\ell_4-\ell_5)^2 \Big\} \nonumber \\[1.0em]
 = & \ \frac{0.00347222}{\ep^8}-\frac{0.000000002(63)}{\ep^7}+\frac{0.0114231(13)}{\ep^6}+\frac{1.163106(20)}{\ep^5} +\frac{14.04762(26)}{\ep^4} \nonumber\\ 
 & \ +\frac{109.8742(34)}{\ep^3}+\frac{647.669(44)}{\ep^2}+\frac{3530.846\pm 1.921}{\ep}\, , \label{eq:2512lines}
\end{align}
\begin{align}
I_{5,{\rm PSLQ}}^{(25)}  =
& \frac{1}{288\ep^8}  + \frac{\zeta_2}{144\ep^6}  + \frac{209\zeta_3}{216\ep^5} + \frac{623\zeta_4}{48\ep^4} +{\cal O}(\ep^{-3}) \,.
\end{align}


\subsection*{Topology~26}
\label{sec:top2612}

\begin{align}
I_{6}^{(26)}  =& \begin{tabular}{c}{\includegraphics[height=1.6cm]{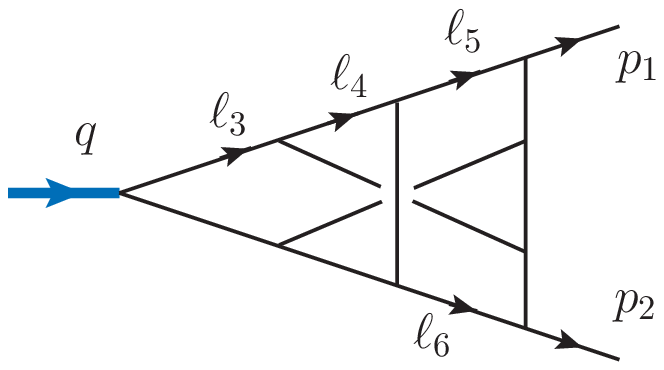}}\end{tabular}   \hskip -.5cm
\times  \, \Big\{ [(\ell_3-\ell_4-\ell_5)^2-(\ell_3-\ell_4-p_1)^2-(\ell_6-p_2)^2-\ell_5^2]  \nonumber \\[1.0em]
                &  \hspace*{25pt}\times [\ell_5^2-\ell_4^2-\ell_6^2+(\ell_4-\ell_6)^2] +4\, \ell_5^2 \, (\ell_6-p_2)^2 + (\ell_4-\ell_5)^2 \, (\ell_3-\ell_4+\ell_6-p_2)^2 \Big\} \nonumber \\[1.0em]
                = & \  -\frac{0.0434028}{\ep^8}-\frac{0.00000002(59)}{\ep^7}+\frac{1.787720(6)}{\ep^6}+\frac{6.90626(7)}{\ep^5} -\frac{13.7958(8)}{\ep^4} \nonumber\\
& \ -\frac{225.841(9)}{\ep^3}-\frac{864.635(105)}{\ep^2} -\frac{9.144 \pm 2.933}{\ep}\, , \label{eq:2612linesA}
\end{align}
\begin{align}
I_{6,{\rm PSLQ}}^{(26)}  =
& -\frac{25}{576\ep^8}  + \frac{313\zeta_2}{288\ep^6}  + \frac{1241\zeta_3}{216\ep^5} - \frac{3671\zeta_4}{288\ep^4} +{\cal O}(\ep^{-3}) \,.
\end{align}


\begin{align}
I_{7}^{(26)}  =& \begin{tabular}{c}{\includegraphics[height=1.6cm]{top26}}\end{tabular}   \hskip -.5cm
\times  \, \Big\{ 4\, [(\ell_4-\ell_5) (\ell_3-\ell_4+\ell_5-p_1)] \, [(\ell_4-\ell_6) (\ell_3-\ell_4+\ell_6-p_2)] \nonumber \\[1.0em]
                &  \hspace*{25pt} - \ell_5^2 \, (\ell_6-p_2)^2 - 4\, (\ell_4-\ell_5)^2 \, (\ell_3-\ell_4+\ell_6-p_2)^2 - \ell_6^2 \, (\ell_5-p_1)^2\nonumber \\[1.0em]
		&  \hspace*{25pt} - (\ell_3-\ell_4)^2 \, (\ell_5+\ell_6-\ell_4)^2 - \ell_4^2 \, (\ell_3-\ell_4+\ell_5+\ell_6-p_1-p_2)^2\Big\} \nonumber \\[1.0em]
               = & \ \frac{0.00347222}{\ep^8}-\frac{0.0000000013}{\ep^7}+\frac{0.0114231(17)}{\ep^6}+\frac{1.16310(3)}{\ep^5} +\frac{2.90880(35)}{\ep^4} \nonumber\\
& \ -\frac{12.2720(43)}{\ep^3}+\frac{29.708(57)}{\ep^2} +\frac{3185.60 \pm 2.63}{\ep}\, , \label{eq:2612linesB}
\end{align}
\begin{align}
I_{7,{\rm PSLQ}}^{(26)}  =
& \frac{1}{288\ep^8}  + \frac{\zeta_2}{144\ep^6}  + \frac{209\zeta_3}{216\ep^5} + \frac{43\zeta_4}{16\ep^4} +{\cal O}(\ep^{-3}) \,.
\end{align}


\subsection*{Topology~27}
\label{sec:top2712}

\begin{align}
I_{8}^{(27)}  =& \begin{tabular}{c}{\includegraphics[height=1.6cm]{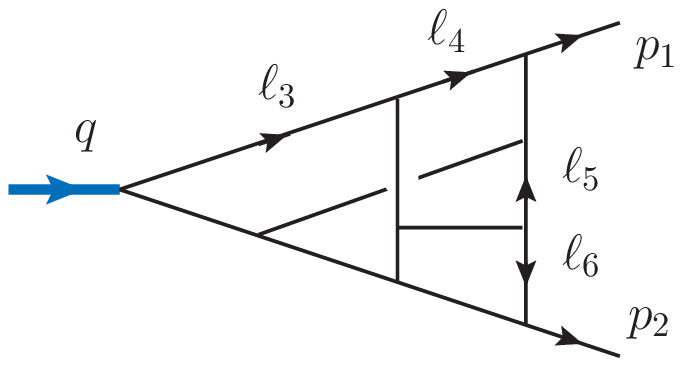}}\end{tabular}   \hskip -.5cm
\times  \, \frac{1}{2} \left[\ell_3^2 - \ell_4^2 - (\ell_4-\ell_3-p_1)^2\right] \, \left[(\ell_3-\ell_4-\ell_5)^2 + (\ell_5+p_2)^2\right] \nonumber \\[1.0em]
 = & \ -\frac{0.015625}{\ep^8}+\frac{0.00000001(14)}{\ep^7}+\frac{0.3426942(17)}{\ep^6}+\frac{1.377357(20)}{\ep^5} \nonumber\\
& \ +\frac{0.41430(24)}{\ep^4} -\frac{18.1972(33)}{\ep^3}-\frac{155.896(52)}{\ep^2} - \frac{1304.61(93)}{\ep} \, , \label{eq:2712lines}
\end{align}
\begin{align}
I_{8,{\rm PSLQ}}^{(27)}
& = -\frac{1}{64\ep^8}  + \frac{5\zeta_2}{24\ep^6}  + \frac{55\zeta_3}{48\ep^5} + \frac{49\zeta_4}{128\ep^4} +{\cal O}(\ep^{-3})\,.
\end{align}


\subsection*{Topology~28}
\label{sec:top2812}

\begin{align}
I_{9}^{(28)}  =& \begin{tabular}{c}{\includegraphics[height=1.6cm]{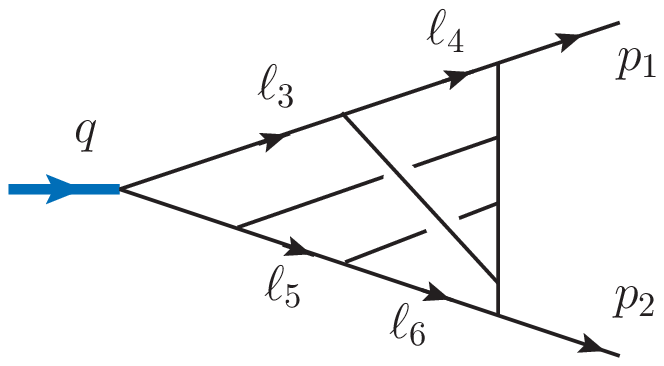}}\end{tabular}   \hskip -.5cm
\times  \, (\ell_3 - \ell_4 - p_2)^2 \, \left[ (\ell_3-\ell_4)^2 - (\ell_3-p_1)^2\right] \nonumber \\[1.0em]
 = & \ -\frac{0.0104167}{\ep^8}+\frac{0.000000002(253)}{\ep^7}+\frac{0.554023(5)}{\ep^6}+\frac{2.26219(5)}{\ep^5}  \nonumber\\
& \ -\frac{3.56367(64)}{\ep^4} -\frac{60.6800(73)}{\ep^3}-\frac{182.180(84)}{\ep^2}+\frac{395.094(952)}{\ep} \, , \label{eq:2812lines}
\end{align}
\begin{align}
I_{9,{\rm PSLQ}}^{(28)}
& = -\frac{1}{96\ep^8}  + \frac{97\zeta_2}{288\ep^6}  + \frac{271\zeta_3}{144\ep^5} - \frac{3793\zeta_4}{1152\ep^4} +{\cal O}(\ep^{-3}) \,.
\end{align}


\subsection*{Topology~29}
\label{sec:top2912}

\begin{align}
I_{10}^{(29)}  =& \begin{tabular}{c}{\includegraphics[height=1.6cm]{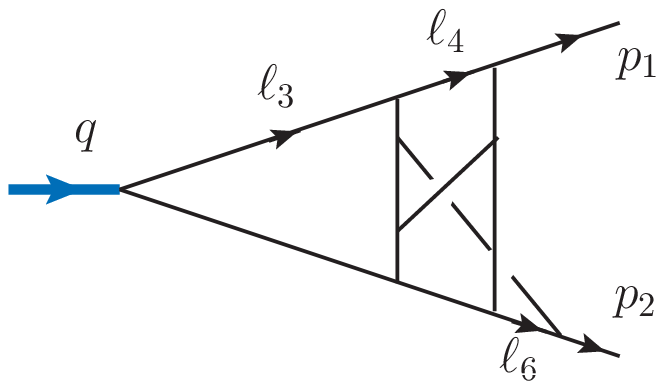}}\end{tabular}   \hskip -.5cm
\times   \, \frac{1}{2} \left[\ell_3^2 - \ell_4^2 - (\ell_4-\ell_3-p_1)^2\right] \, \left[\ell_6 \cdot (\ell_6 - \ell_4 + \ell_3 - p_2)\right] \nonumber \\[1.0em]
= &\ -\frac{0.000868056}{\ep^8}+\frac{0.0000000005}{\ep^7}-\frac{0.00285575(22)}{\ep^6}-\frac{0.0090438(31)}{\ep^5} +\frac{0.714516(37)}{\ep^4} \nonumber\\
&\ +\frac{10.2737(4)}{\ep^3}+\frac{76.5178(52)}{\ep^2}+\frac{370.489(160)}{\ep} \, , \label{eq:2912lines}
\end{align}
\begin{align}
I_{10,{\rm PSLQ}}^{(29)}  =
& -\frac{1}{1152\ep^8}  - \frac{\zeta_2}{576\ep^6}  - \frac{13\zeta_3}{1728\ep^5} + \frac{169\zeta_4}{256\ep^4} +{\cal O}(\ep^{-3}) \,.
\end{align}


\subsection*{Topology~30}
\label{sec:top3012}

\begin{align}
I_{11}^{(30)}  =& \begin{tabular}{c}{\includegraphics[height=1.6cm]{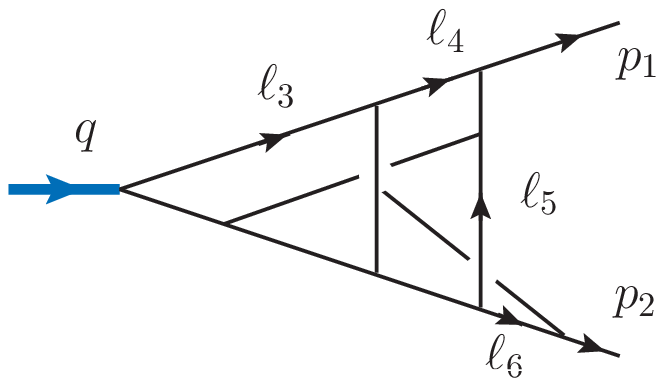}}\end{tabular}   \hskip -.5cm
\times \, (\ell_3-\ell_4-p_2)^2 \, [(p_1-\ell_4)^2+(\ell_3-\ell_4)^2-(\ell_3-p_1)^2] \nonumber \\[1.0em]
                =& \frac{0.00347222}{\ep^8}-\frac{0.05140419}{\ep^6}-\frac{0.2601674}{\ep^5} -\frac{1.5145009}{\ep^4}  \nonumber\\
		&\ - \frac{17.34721164(4)}{\ep^3}-\frac{133.31287(3)}{\ep^2}-\frac{671.48(24)}{\ep} \, . \label{eq:3012lines}
\end{align}
This result was obtained with MB. FIESTA performs poorly in this topology.
\begin{align}
I_{11,{\rm PSLQ}}^{(30)}  =
& \frac{1}{288\ep^8}  - \frac{\zeta_2}{32\ep^6}  - \frac{187\zeta_3}{864\ep^5} - \frac{403\zeta_4}{288\ep^4} +{\cal O}(\ep^{-3}) \,.
\end{align}


\subsection{UT integrals with 11 lines}
\label{sec:utints11}


\subsection*{Topology~27}
\label{sec:top2711}

\begin{align}
I_{12}^{(27)}  =& \begin{tabular}{c}{\includegraphics[height=1.6cm]{top27}}\end{tabular}   \hskip -.5cm
\times \, \frac{1}{2} \, (\ell_3-\ell_4)^2 \, \left[2 \, (\ell_4-p_2)^2 + (\ell_6-p_1)^2 \right.  \nonumber \\ &\hskip 6cm \left.- (\ell_4-\ell_6)^2 - \ell_4^2 + \ell_5^2 + 2 \, (p_1+p_2)^2\right] \nonumber \\[1.0em]
 =&\ \frac{0.0303819}{\ep^8}-\frac{0.00000002(87)}{\ep^7}-\frac{0.625418(2)}{\ep^6}-\frac{2.824274(22)}{\ep^5} -\frac{7.64568(40)}{\ep^4} \nonumber\\
&\ -\frac{22.7148(82)}{\ep^3}+\frac{0.160(47)}{\ep^2} +\frac{1354.58(99)}{\ep}\, . \label{eq:2711lines}
\end{align}
This result is obtained by combining FIESTA and MB results.
\begin{align}
I_{12,{\rm PSLQ}}^{(27)}
& = \frac{35}{1152\ep^8}  - \frac{73\zeta_2}{192\ep^6}  - \frac{1015\zeta_3}{432\ep^5} - \frac{4069\zeta_4}{576\ep^4} +{\cal O}(\ep^{-3})  \,.
\end{align}


\subsection*{Topology~28}
\label{sec:top2811}

\begin{align}
I_{13}^{(28)}  =& \begin{tabular}{c}{\includegraphics[height=1.6cm]{top28}}\end{tabular}   \hskip -.5cm
\times \, \frac{1}{2} \, (\ell_3-\ell_4)^2 \, \left[2 \, (\ell_3-\ell_4-p_2)^2 + (\ell_6-p_1)^2 - (\ell_4-\ell_6)^2 + \ell_4^2\right] \nonumber \\[1.0em]
= & \ -\frac{0.0112847}{\ep^8}+\frac{0.000000001(95)}{\ep^7}+\frac{0.299858(2)}{\ep^6}+\frac{0.848669(24)}{\ep^5} +\frac{0.86617(24)}{\ep^4}  \nonumber\\
&\ +\frac{10.3884(22)}{\ep^3}+\frac{107.036(19)}{\ep^2} +\frac{184.841\pm 1.038}{\ep}\, , \label{eq:2811lines}
\end{align}
\begin{align}
I_{13,{\rm PSLQ}}^{(28)}  =
& -\frac{13}{1152\ep^8}  + \frac{35\zeta_2}{192\ep^6}  + \frac{305\zeta_3}{432\ep^5} + \frac{461\zeta_4}{576\ep^4} +{\cal O}(\ep^{-3}) \,.
\end{align}


\subsection*{Topology~29}
\label{sec:top2911}

\begin{align}
I_{14}^{(29)}  =& \begin{tabular}{c}{\includegraphics[height=1.6cm]{top29}}\end{tabular}   \hskip -.5cm
\times \, (\ell_4-p_1)^2 \, \left[(\ell_3-\ell_4+\ell_6)^2 + (\ell_6-p_2)^2 - \ell_6^2\right] \nonumber \\[1.0em]
 =& \ -\frac{0.03756430(3)}{\ep^5}  +\frac{0.0761009(7)}{\ep^4}-\frac{0.258654(13)}{\ep^3}+\frac{1.68939(17)}{\ep^2} +\frac{81.404(26)}{\ep} \, , \label{eq:2911linesA}
\end{align}
\begin{align}
I_{14,{\rm PSLQ}}^{(29)}  =
& -\frac{\zeta_3}{32\ep^5} + \frac{9\zeta_4}{128\ep^4} +{\cal O}(\ep^{-3}) \,.
\end{align}


\begin{align}
I_{15}^{(29)}  =& \begin{tabular}{c}{\includegraphics[height=1.6cm]{top29}}\end{tabular}   \hskip -.5cm
\times \, \frac{1}{2} \, (\ell_3-p_1-p_2)^2 \, \left[(\ell_4-\ell_6)^2 - (\ell_4-p_2)^2- (\ell_6-p_1)^2 - (p_1+p_2)^2 \right] \nonumber \\[1.0em]
 =&\ -\frac{0.0555556}{\ep^8}+\frac{0.000000003(192)}{\ep^7}+\frac{1.816280(3)}{\ep^6}+\frac{7.29303(4)}{\ep^5} -\frac{2.67392(45)}{\ep^4}  \nonumber\\
&\ -\frac{134.463(5)}{\ep^3}-\frac{642.326(56)}{\ep^2} -\frac{1530.04(97)}{\ep}\, , \label{eq:2911linesB}
\end{align}
\begin{align}
I_{15,{\rm PSLQ}}^{(29)}
& = -\frac{1}{18\ep^8}  + \frac{53\zeta_2}{48\ep^6}+\frac{2621\zeta_3}{432\ep^5} - \frac{1423\zeta_4}{576\ep^4} +{\cal O}(\ep^{-3})  \,.
\end{align}


\subsection*{Topology~30}
\label{sec:top3011}

\begin{align}
I_{16}^{(30)}  =& \begin{tabular}{c}{\includegraphics[height=1.6cm]{top30}}\end{tabular}   \hskip -.5cm
\times  \, (\ell_3-p_1-p_2)^2 \, (\ell_5+p_2)^2 \nonumber \\[1.0em]
 = & \ \frac{0.036458333}{\ep^8}-\frac{0.5997155452(1)}{\ep^6}-\frac{2.2622043108(1)}{\ep^5} -\frac{0.828653725(2)}{\ep^4}\nonumber\\
&\ +\frac{19.82059(28)}{\ep^3}+\frac{94.8794(349)}{\ep^2} +\frac{232.242(541)}{\ep}\, , \label{eq:3011linesA}
\end{align}
\begin{align}
I_{16,{\rm PSLQ}}^{(30)}  
& = \frac{7}{192\ep^8}  - \frac{35\zeta_2}{96\ep^6}-\frac{271\zeta_3}{144\ep^5} - \frac{49\zeta_4}{64\ep^4}+{\cal O}(\ep^{-3})  \,.
\end{align}


\begin{align}
I_{17}^{(30)}  =& \begin{tabular}{c}{\includegraphics[height=1.6cm]{top30}}\end{tabular}   \hskip -.5cm
\times  \, \frac{1}{2} \, (\ell_4-p_1)^2 \, \left[2\, (\ell_5+p_2)^2 - (\ell_5+p_2+\ell_4-\ell_3)^2 \right] \nonumber \\[1.0em]
= & \ -\frac{0.03756430(4)}{\ep^5}  +\frac{0.1042870(7)}{\ep^4}+\frac{1.64150(1)}{\ep^3}+\frac{8.56434(14)}{\ep^2} +\frac{35.4679(216)}{\ep}\, , \label{eq:3011linesB}
\end{align}
\begin{align}
I_{17,{\rm PSLQ}}^{(30)}  =
& -\frac{\zeta_3}{32\ep^5} + \frac{37\zeta_4}{384\ep^4} +{\cal O}(\ep^{-3}) \,.
\end{align}


\begin{align}
I_{18}^{(30)}  =& \begin{tabular}{c}{\includegraphics[height=1.6cm]{top30}}\end{tabular}   \hskip -.5cm
\times  \, \frac{1}{2} \, (\ell_3-\ell_4)^2 \, \left[2\, (\ell_6-\ell_4+p_1)^2 - 3 \, \ell_6^2 \right] \nonumber \\[1.0em]
 = & \ \frac{0.00347222}{\ep^8}-\frac{0.0000000002(316)}{\ep^7}+\frac{0.0628273(8)}{\ep^6}-\frac{0.001391(11)}{\ep^5} \nonumber\\
& \  -\frac{4.07561(14)}{\ep^4} -\frac{35.6750(18)}{\ep^3}-\frac{211.233(25)}{\ep^2} -\frac{1162.74(39)}{\ep}\, , \label{eq:3011linesC}
\end{align}
\begin{align}
I_{18,{\rm PSLQ}}^{(30)}  =
& \frac{1}{288\ep^8}  + \frac{11\zeta_2}{288\ep^6}-\frac{\zeta_3}{864\ep^5} - \frac{241\zeta_4}{64\ep^4}  +{\cal O}(\ep^{-3}) \,.
\end{align}


\subsection{UT integrals with 10 lines}
\label{sec:utints10}


\subsection*{Topology~22}
\label{sec:top2210}

\begin{align}
I_{19}^{(22)}  =& \begin{tabular}{c}{\includegraphics[height=1.6cm]{top22}}\end{tabular}  \hskip -.5cm
\times  \, (\ell_3-\ell_4)^2 \,  (p_1-\ell_3+\ell_6)^2 \nonumber \\[1.0em]
                  = & \ \frac{0.00173611111}{\ep^8}+\frac{0.165635722(1)}{\ep^6}+\frac{0.74850303(1)}{\ep^5} +\frac{4.1564218626(4)}{\ep^4} \nonumber\\
&\ +\frac{36.5261(6)}{\ep^3}+\frac{315.366(13)}{\ep^2}+\frac{2180.03(19)}{\ep} \, ,\label{eq:2210linesA}
\end{align}
\begin{align}
I_{19,{\rm PSLQ}}^{(22)}  =
& \frac{1}{576\ep^8} + \frac{29\zeta_2}{288\ep^6}  + \frac{269\zeta_3}{432\ep^5} + \frac{553\zeta_4}{144\ep^4} +{\cal O}(\ep^{-3}) \,.
\end{align}


\begin{align}
I_{20}^{(22)}  =& \begin{tabular}{c}{\includegraphics[height=1.6cm]{top22}}\end{tabular} \hskip -.5cm
\times  \, \ell_6^2 \,  (p_1-\ell_4)^2 =\frac{1.34678628(2)}{\ep^2} - \frac{6.89677(9)}{\ep} \, .\label{eq:2210linesB}
\end{align}


\subsection*{Topology~24}
\label{sec:top2410}

\begin{align}
I_{21}^{(24)}  =& \begin{tabular}{c}{\includegraphics[height=1.6cm]{top24}}\end{tabular}   \hskip -.5cm
\times  \, (p_1 - \ell_3-\ell_5)^2 \,  (\ell_3-p_1-p_2)^2 \nonumber \\[1.0em]
                 =& \ \frac{0.00868056}{\ep^8}-\frac{0.0000000009(316)}{\ep^7}+\frac{0.211328(1)}{\ep^6}+\frac{0.637202(14)}{\ep^5} \nonumber\\
& \ -\frac{4.06623(11)}{\ep^4} -\frac{48.3099(8)}{\ep^3}-\frac{242.796(6)}{\ep^2}-\frac{819.895(471)}{\ep} \, ,\label{eq:2410linesA}
\end{align}
\begin{align}
I_{21,{\rm PSLQ}}^{(24)}  =
& \frac{5}{576\ep^8} + \frac{37\zeta_2}{288\ep^6}  + \frac{229\zeta_3}{432\ep^5} - \frac{541\zeta_4}{144\ep^4} +{\cal O}(\ep^{-3}) \,.
\end{align}


\begin{align}
I_{22}^{(24)}  =& \begin{tabular}{c}{\includegraphics[height=1.6cm]{top24}}\end{tabular}   \hskip -.5cm
\times  \, \ell_5^2 \,  (\ell_3-p_1-p_2)^2 \nonumber \\[1.0em]
                 =& \ \frac{0.00694444}{\ep^8}-\frac{0.000000001(32)}{\ep^7}-\frac{0.0913852(12)}{\ep^6}-\frac{1.46362(1)}{\ep^5} -\frac{11.7515(1)}{\ep^4} \nonumber\\
& \ -\frac{66.0916(9)}{\ep^3}-\frac{236.916(8)}{\ep^2}-\frac{59.966(634)}{\ep} \, ,\label{eq:2410linesB}
\end{align}
\begin{align}
I_{22,{\rm PSLQ}}^{(24)}  =
& \frac{1}{144\ep^8} - \frac{\zeta_2}{18\ep^6}  - \frac{263\zeta_3}{216\ep^5} - \frac{3127\zeta_4}{288\ep^4}+{\cal O}(\ep^{-3}) \,.
\end{align}


\subsection*{Topology~28}
\label{sec:top2810}

\begin{align}
I_{23}^{(28)}  =& \begin{tabular}{c}{\includegraphics[height=1.6cm]{top28}}\end{tabular}   \hskip -.5cm
\times \, (\ell_4 - p_1)^2 \,  (\ell_3-\ell_4+\ell_5-p_2)^2 \nonumber \\[1.0em]
                 &= \frac{0.6764520(3)}{\ep^4}+\frac{2.779140(4)}{\ep^3}+\frac{3.67317(5)}{\ep^2}-\frac{26.2248(38)}{\ep}  \, , \label{eq:2810lines}
\end{align}
\begin{align}
I_{23,{\rm PSLQ}}^{(28)}  =
&  \frac{5\zeta_4}{8\ep^4} +{\cal O}(\ep^{-3}) \,.
\end{align}

\section{Basis of propagators and numerators}
\label{app:basis}

This appendix contains the basis of 12 propagators and 6 irreducible numerators, which are used in section \ref{sec:PLUT}. The numbering of the equations corresponds to the topologies in figure~\ref{fig:Ptops}~--~\ref{fig:NPtops}. In each case, the first twelve entries parametrise the twelve propagators of the respective integral and the last six entries the chosen numerators. We have defined $q = p_1 + p_2$.

\abovedisplayskip=1pt
\belowdisplayskip=1pt
\begin{multline}
\{   l_6  , \, l_5  , \, l_4  , \, l_3  , \,  l_6-l_5   , \,  l_4-l_3   , \,  p_1-l_6  , \, 
    l_5-l_4   , \,  -l_6+q   , \,  -l_4+q   , \,  -l_3+q  , \, 
    -l_5+q   , \\ 
    l_3-l_5   , \,  l_3-l_6   , \,  l_4-l_6   , \,  l_3-p_2  , \, 
    l_4-p_2   , \,  l_5-p_2 \}  \,,
\end{multline}
\begin{multline}
\{   l_6  , \, l_5  , \, l_4  , \, l_3  , \,  l_5-l_4   , \,  l_6-l_5   , \,  p_1-l_6  , \, 
    l_3-l_4   , \,  -l_3+q   , \,  -l_6+q   , \,  -l_5+q  , \, 
    -l_3+l_4-l_5+q   , \\  
     l_3-l_5   , \,  l_3-l_6   , \,  l_4-l_6  , \, 
    l_4-p_1   , \,  l_5-p_2   , \,  l_3-p_2    \}  \,,
\end{multline}
\begin{multline}
\{   l_6  , \, l_5  , \, l_4  , \, l_3  , \,  l_4-l_3   , \,  l_4-l_5   , \,  l_6-l_5  , \, 
    p_1-l_6   , \,  -l_4+q   , \,  -l_6+q   , \,  -l_3+q  , \, 
    -l_4+l_5-l_6+q   ,  \\  
     l_5-p_1   , \,  l_3-p_2   , \,  l_4-p_2  , \, 
    l_5-p_2   , \,  l_3-l_5   , \,  l_3-l_6    \}  \,,
\end{multline}
\begin{multline}
\{   l_6  , \, l_5  , \, l_4  , \, l_3  , \,  p_1-l_6   , \,  l_4-l_3   , \,  l_5-l_6  , \, 
    l_5-l_4   , \,  -l_5+l_6+p_2   , \,  -l_5+q   , \,  -l_3+q  , \, 
    -l_4+q   , \\   
     l_3-l_5   , \,  l_3-l_6   , \,  l_4-l_6   , \,  l_3-p_2  , \, 
    l_4-p_2   , \,  l_6-p_2    \}  \,,
\end{multline}
\begin{multline}
\{   l_6  , \, l_5  , \, l_4  , \, l_3  , \,  l_3-l_4   , \,  l_5-l_4   , \,  p_1-l_6  , \, 
    l_5-l_6   , \,  -l_5+q   , \,  -l_5+l_6+p_2   , \,  -l_3+q  , \, 
    -l_3+l_4-l_5+q   , \\    l_3-p_2   , \,  l_3-l_6   , \,  l_4-l_6  , \, 
    l_4-p_1   , \,  l_4-p_2   , \,  l_6-p_2    \}  \,,
\end{multline}
\begin{multline}
\{   l_6  , \, l_5  , \, l_4  , \, l_3  , \,  p_1-l_5   , \,  l_4-l_5   , \,  l_4-l_3  , \, 
    p_2-l_6   , \,  -l_3+q   , \,  -l_4+q   , \,  -l_4-l_6+q  , \, 
    -l_5-l_6+q   , \\  
    l_3-l_5   , \,  l_3-l_6   , \,  l_4-l_6  , \, 
    l_5-l_6   , \,  l_3-p_2   , \,  l_4-p_2    \}  \,,
\end{multline}
\begin{multline}
\{   l_6  , \, l_5  , \, l_4  , \, l_3  , \,  p_1-l_5   , \,  l_4-l_5   , \,  l_4-l_3  , \, 
    p_2-l_6   , \,  -l_4+q   , \,  -l_4+l_5+l_6   , \,  -l_3+q  , \, 
    -l_5-l_6+q   , \\  
    l_3-l_5   , \,  l_3-l_6   , \,  l_5-l_6  , \, 
    l_3-p_2   , \,  l_4-p_2   , \,  l_5-p_2   \}  \,,
\end{multline}
\begin{multline}
\{   l_6  , \, l_5  , \, l_4  , \, l_3  , \,  p_2-l_6   , \,  l_4-l_5   , \,  l_4-l_3  , \, 
    p_1-l_5   , \,  -l_4+q   , \,  -l_3+q   , \,  -l_4-l_6+q  , \, 
    l_4-l_5+l_6-p_2   , \\    
    l_3-l_5   , \,  l_3-l_6   , \,  l_4-l_6   , \,  l_5-l_6  
   , \,  l_3-p_2   , \,  l_5-p_2    \}  \,,
\end{multline}
\begin{multline}
\{   l_6  , \, l_5  , \, l_4  , \, l_3  , \,  l_4-l_3   , \,  p_1-l_5   , \,  l_4-l_5  , \, 
    -l_5+q   , \,  -l_3+q   , \,  -l_4-l_6+q  , \, 
    l_5+l_6-q   , \,  -l_3-l_6+q   , \\    
    l_3-l_5   , \,  l_3-p_2  , \, 
    l_4-p_2   , \,  l_4-l_6   , \,  l_5-l_6   , \,  l_6-p_2    \}  \,,
\end{multline}
\begin{multline}
\{   l_6  , \, l_5  , \, l_4  , \, l_3  , \,  l_4-l_5   , \,  p_1-l_5   , \,  l_3-l_4  , \, 
    l_3-l_4+l_6   , \,  -l_3+q   , \,  -l_5+q  ,  \, 
    -l_3-l_6+q   , \,  l_3-l_4+l_5+l_6-q   , \\    l_5-l_6  , \, 
    p_2-l_6   , \,  l_3-l_5   , \,  l_4-p_2   , \,  l_3-p_2   , \,  l_4-l_6   \}  \,,
\end{multline}
\begin{multline}
\{   l_6  , \, l_5  , \, l_4  , \, l_3  , \,  l_4-l_5   , \,  p_1-l_5   , \,  l_4-l_3  , \, 
    -l_3+q   , \,  -l_5+q   , \,  l_5+l_6-q  , \,
    -l_3-l_6+q   , \,  l_3-l_4+l_5+l_6-q   , \\    l_3-l_5  , \, 
    l_3-p_2   , \,  l_4-p_1   , \,  l_4-p_2   , \,  l_5-l_6   , \,  l_6-p_2   \}  \,,
\end{multline}
\begin{multline}
\{   l_6  , \, l_5  , \, l_4  , \, l_3  , \,  p_2-l_6   , \,  p_1-l_5   , \,  l_3-l_4  , \, 
    l_5-l_4   , \,  -l_3+q   , \,  -l_4+l_5+l_6   , \,  -l_5-l_6+q  , \,
    -l_3+l_4-l_5-l_6+q   , \\   
    l_3-l_6   , \,  l_5-p_2   , \,  l_3-p_2  , \, 
    l_4-p_1   , \,  l_4-p_2   , \,  l_6-p_1    \}  \,,
\end{multline}
\begin{multline}
\{   l_6  , \, l_5  , \, l_4  , \, l_3  , \,  l_3-l_4   , \,  p_2-l_6   , \,  p_1-l_4  , \, 
    l_6-l_5   , \,  -l_3+q   , \,  -l_4-l_6+q  , \, 
    -l_3-l_5+q   , \,  -l_4-l_5+q   , \\    l_3-l_6   , \,  l_3-p_2  , \, 
    l_4-p_2   , \,  l_5-p_1   , \,  l_5-p_2   , \,  l_6-p_1    \}  \,,
\end{multline}
\begin{multline}
\{   l_6  , \, l_5  , \, l_4  , \, l_3  , \,  p_2-l_6   , \,  p_1-l_4   , \,  l_3-l_4  , \, 
    l_6-l_5   , \,  -l_3+l_4+l_5   , \,  -l_3+q   , \,  -l_4-l_5+q  , \, 
    -l_4-l_6+q   , \\    l_4-l_5   , \,  l_3-l_6   , \,  l_4-l_6  , \, 
    l_3-p_2   , \,  l_4-p_2   , \,  l_5-p_2    \}  \,, 
\end{multline}
\begin{multline}
\{   l_6  , \, l_5  , \, l_4  , \, l_3  , \,  p_2-l_6   , \,  p_1-l_4   , \,  l_3-l_4  , \, 
    l_5-l_6   , \,  -l_3+q   , \,  -l_4-l_5+l_6+p_1  , \, 
    -l_3-l_5+q   , \,  -l_4-l_5+q   , \\    l_3-l_6   , \,  l_3-p_2  , \, 
    l_4-p_2   , \,  l_5-p_1   , \,  l_5-p_2   , \,  l_6-p_1    \}  \,, 
\end{multline}
\begin{multline}
\{   l_6  , \, l_5  , \, l_4  , \, l_3  , \,  p_2-l_6   , \,  p_1-l_4   , \,  l_6-l_5  , \, 
    l_3-l_4   , \,  -l_3+q   , \,  l_3-l_4+l_5-l_6  , \, 
    -l_3-l_5+q   , \,  -l_4-l_6+q   , \\    l_3-l_6   , \,  l_3-p_2  , \, 
    l_4-l_5   , \,  l_4-p_2   , \,  l_5-p_1   , \,  l_6-p_1     \}  \,,
\end{multline}
\begin{multline}
\{   l_6  , \, l_5  , \, l_4  , \, l_3  , \,  p_2-l_6   , \,  l_5-l_4   , \,  p_1-l_5  , \, 
    l_4-l_3   , \,  -l_3+q   , \,  -l_5-l_6+q  , \, 
    -l_3+l_4-l_5+q   , \,  -l_3+l_4-l_5-l_6+q   , \\    
    l_3-p_2  
   , \,  l_4-l_6   , \,  l_5-p_2   , \,  l_4-p_1   , \,  l_4-p_2   , \,  l_6-p_1    \}  \,,
\end{multline}
\begin{multline}
\{   l_6  , \, l_5  , \, l_4  , \, l_3  , \,  p_1-l_5   , \,  p_2-l_6   , \,  l_3-l_4  , \, 
    l_5-l_4   , \,  l_3-l_4-l_6   , \,  -l_3+q  , \,
    -l_3+l_4-l_5+q   , \,  -l_3+l_4-l_5+l_6+p_1   , \\    l_3-p_2  , \, 
    l_3-l_5   , \,  l_4-l_6   , \,  l_4-p_2   , \,  l_5-p_2   , \,  l_6-p_1    \}  \,,
\end{multline}
\begin{multline}
\{   l_6  , \, l_5  , \, l_4  , \, l_3  , \,  p_2-l_6   , \,  p_1-l_5   , \,  l_4-l_5  , \, 
    l_3-l_4   , \,  -l_4+l_5+l_6   , \,  -l_3+q   , \,  -l_3+l_5+l_6  , \,
    -l_5-l_6+q   , \\    
    l_3-l_6   , \,  l_5-l_6   , \,  l_4-p_1  , \, 
    l_3-p_2   , \,  l_4-p_2   , \,  l_5-p_2     \}  \,,
\end{multline}
\begin{multline}
\{   l_6  , \, l_5  , \, l_4  , \, l_3  , \,  p_2-l_6   , \,  l_4-l_3   , \,  p_1-l_5  , \, 
    l_5-l_4   , \,  -l_3+l_5+l_6   , \,  -l_3+q   , \,  -l_3+l_4+l_6  , \,
    -l_5-l_6+q   , \\   
    l_4-l_6   , \,  l_5-l_6   , \,  l_4-p_1  , \, 
    l_3-p_2   , \,  l_4-p_2   , \,  l_5-p_2     \}  \,,
\end{multline}
\begin{multline}
\{   l_6  , \, l_5  , \, l_4  , \, l_3  , \,  p_1-l_4   , \,  l_3-l_4   , \,  p_2-l_6  , \, 
    l_5+l_6   , \,  -l_4-l_5+p_1   , \,  -l_3+q   , \,  -l_3+l_6+p_1  , \,
    -l_3-l_5+p_1   , \\    
    l_3-p_2   , \,  l_4-p_2   , \,  l_5-p_2   , \,  l_6-p_1  , \, 
    l_4-l_6   , \,  l_3-l_5     \}  \,,
\end{multline}
\begin{multline}
\{   l_6  , \, l_5  , \, l_4  , \, l_3  , \,  l_3-l_4   , \,  l_5+l_6   , \,  p_2-l_6  , \, 
    p_1-l_4   , \,  -l_3+l_6+p_1   , \,  -l_3+q   , \,  -l_4-l_5+p_1  , \,
    -l_3+l_4+l_5+l_6   , \\    
    l_3-l_5   , \,  l_3-p_2   , \,  l_4-p_2  , \, 
    l_5-p_1   , \,  l_5-p_2   , \,  l_6-p_1     \}  \,,
\end{multline}
\begin{multline}
\{   l_6  , \, l_5  , \, l_4  , \, l_3  , \,  l_6-p_2   , \,  p_1-l_4   , \,  l_6-l_5  , \, 
    l_3-l_4   , \,  -l_3+q   , \,  l_3-l_4-l_5   , \,  -l_3+l_6+p_1  , \,
    -l_3+l_5+p_1   , \\  
    l_3-p_2   , \,  l_4-p_2   , \,  l_5-p_1   , \,  l_6-p_1  , \, 
    l_5-p_2   , \,  l_4-l_6     \}  \,,
\end{multline}
\begin{multline}
\{   l_6  , \, l_5  , \, l_4  , \, l_3  , \,  p_2-l_6   , \,  l_5+l_6   , \,  l_3-l_4  , \, 
    p_1-l_4   , \,  -l_3-l_5+p_1   , \,  -l_4-l_5+p_1   , \,  -l_3+q  , \,
    -l_3-l_5-l_6+q   , \\    
    l_3-p_2   , \,  l_4-l_6   , \,  l_4-p_2  , \, 
    l_5-p_1   , \,  l_5-p_2   , \,  l_6-p_1     \}  \,,
\end{multline}
\begin{multline}
\{   l_6  , \, l_5  , \, l_4  , \, l_3  , \,  p_1-l_5   , \,  l_5-l_4   , \,  p_2-l_6  , \, 
    l_3-l_4   , \,  -l_4+l_5+l_6   , \,  -l_3+q   , \,  -l_3+l_4-l_5+p_1  , \,
    -l_3+l_4-l_5-l_6+q   , \\    
    l_3-l_5   , \,  l_3-l_6   , \,  l_5-l_6  , \, 
    l_4-p_1   , \,  l_4-p_2   , \,  l_5-p_2     \}  \,,
\end{multline}
\begin{multline}
\{   l_6  , \, l_5  , \, l_4  , \, l_3  , \,  l_5-l_4   , \,  l_3-l_4   , \,  p_1-l_5  , \, 
    p_2-l_6   , \,  -l_4+l_5+l_6   , \,  -l_3+q   , \,  l_3-l_4+l_6-p_2  , \, -l_3+l_4-l_5-l_6+q   , \\  
     l_3-l_5   , \,  l_3-l_6   , \,  l_5-l_6  , \, 
    l_4-p_1   , \,  l_4-p_2   , \,  l_5-p_2     \}  \,,
\end{multline}
\begin{multline}
\{   l_6  , \, l_5  , \, l_4  , \, l_3  , \,  l_5+l_6   , \,  p_2-l_6   , \,  p_1-l_4  , \, 
    l_3-l_4   , \,  -l_4-l_5+p_1   , \,  -l_3+q   , \,  -l_3+l_4+l_5+p_2  , \,
    -l_3+l_4+l_5+l_6   , \\    
    l_4-l_6   , \,  l_3-p_2   , \,  l_4-p_2  , \, 
    l_5-p_1   , \,  l_5-p_2   , \,  l_6-p_1     \}  \,,
\end{multline}
\begin{multline}
\{   l_6  , \, l_5  , \, l_4  , \, l_3  , \,  l_3-l_4   , \,  p_1-l_4   , \,  p_2-l_6  , \, 
    l_5-l_6   , \,  -l_3+q   , \,  l_3-l_4+l_5-p_2  , \,
    -l_3-l_5+q   , \,  -l_3+l_4-l_6+p_2   , \\    
    l_3-p_2   , \,  l_4-l_5  , \, 
    l_4-l_6   , \,  l_4-p_2   , \,  l_5-p_2   , \,  l_6-p_1     \}  \,,
\end{multline}
\begin{multline}
\{   l_6  , \, l_5  , \, l_4  , \, l_3  , \,  p_1-l_4   , \,  l_4-l_3   , \,  p_2-l_6  , \, 
    l_5+l_6   , \,  -l_4-l_5+p_1   , \,  -l_3+q   , \,  -l_3+l_4-l_6+p_2  , \, 
    -l_3-l_5-l_6+q   , \\    
    l_3-p_2   , \,  l_4-l_6   , \,  l_4-p_2  , \, 
    l_5-p_1   , \,  l_5-p_2   , \,  l_6-p_1     \}  \,,
\end{multline}
\begin{multline}
\{   l_6  , \, l_5  , \, l_4  , \, l_3  , \,  l_3-l_4   , \,  l_5+l_6   , \,  p_2-l_6  , \, 
    p_1-l_4   , \,  -l_3+q   , \,  -l_4-l_5+p_1   , \,  -l_3+l_4+l_5+p_2  , \,
    -l_3+l_4-l_6+p_2   , \\    
    l_3-p_2   , \,  l_4-l_6   , \,  l_4-p_2  , \, 
    l_5-p_1   , \,  l_5-p_2   , \,  l_6-p_1     \}  \,.
\end{multline}
\abovedisplayskip=10pt
\belowdisplayskip=10pt

\bibliographystyle{JHEP}


\providecommand{\href}[2]{#2}\begingroup\raggedright\endgroup

\end{document}